\documentclass{LMCS}

\def\dOi{10(3:17)2014}
\lmcsheading%
{\dOi}
{1--29}
{}
{}
{Jan.~\phantom06, 2014}
{Sep.~10, 2014}
{}

\ACMCCS{[{\bf Mathematics of computing}]: Probability and
  statistics---Stochastic processes---Markov processes; [{\bf Theory of computation}]: Models of computation; Formal
  languages and automata theory}


\usepackage{amssymb}
\usepackage{amsthm}
\usepackage{afterpage}
\usepackage{thmtools,thm-restate}
\usepackage{amsmath}
\usepackage{latexsym}
\usepackage{color}
\usepackage{ifthen}
\usepackage{multirow}
\usepackage{arrows}                  
\usepackage{rotating}
\usepackage{hyperref}
\usepackage{subfigure}
\usepackage{stmaryrd}

\usepackage{tikz}
\usetikzlibrary{automata,fit}
\usetikzlibrary{decorations.pathmorphing}
\tikzstyle{every picture}=[thick]
\tikzstyle{every loop}=[->]
\tikzstyle{every scope}=[>=latex]
\tikzstyle{dot}=[circle,thick,minimum size=0.5mm,fill=black, inner sep=1pt]
\tikzstyle{every state}=[draw=black,line width=.5pt,fill=white,minimum size=10pt,initial text=]

\usetikzlibrary{shapes}
\newlength{\arclength}
\usepackage{dsfont}      
\usepackage{mathtools}   
\usepackage{mathpartir}  
\usepackage{url}
\usepackage{xspace}

\DeclareMathAlphabet{\mathitbf}{OML}{cmm}{b}{it}

\newcommand\distr{\ensuremath{\mathsf{Distr}}}
\newcommand\subdistr{\ensuremath{\mathsf{Subdistr}}}
\newcommand\spt{\ensuremath{\mathsf{supp}}}

\newcommand{\dirac}[1]{\mathds{1}_{#1}}

\renewcommand{\prob}{\mathbb{P}}

\newcommand{\union}{\mathrel{\cup}}

\newcommand{\qdot}{\; . \;}

 \newcommand{\mtrans}[1]{\stackrel{\smash{\lower1pt\hbox{\scriptsize $#1$}\,}}{\lower1pt\hbox{$\rightsquigarrow$}}}

\newcommand{\itrans}[1]{\xhookrightarrow{\smash{\lower1pt\hbox{\scriptsize $\,\smash{#1}\,$}\vphantom{X}}}}

\newcommand{\trans}[1]{\xrightarrow{\smash{#1}}}

\makeatletter 
\providecommand\dotsum{\mathpalette\@dotted\sum \vphantom{\sum}} 
\def\@dotted#1#2{\ooalign{\hfil$#1 \bullet $\hfil\cr\hfil$#1#2$\hfil\cr}} 
\makeatother

\let\ifTIKZ\iffalse
\let\ifTIKZ\iftrue

 \usetikzlibrary{decorations.markings}
 \usetikzlibrary{shapes.geometric}

 \pgfdeclarelayer{edgelayer}
 \pgfdeclarelayer{nodelayer}
 \pgfsetlayers{main,edgelayer,nodelayer}

\usetikzlibrary{arrows,shapes,petri,positioning,calc}
\ifTIKZ
\usetikzlibrary{decorations.text,decorations.pathmorphing}
\fi
\tikzstyle{place}=[circle,thick,draw=black!100,fill=black!0,minimum
size=3mm] 
\tikzstyle{stoTH}=[rectangle,thick, minimum width=1.5mm, minimum
height=5.5mm,thick,draw=black!100, fill=black!0,inner sep=0pt]
\tikzstyle{immTH}=[rectangle, minimum width=1mm, minimum
height=6mm,thin,draw=black!100, fill=black!100,inner sep=0pt]

\tikzstyle{aLabel}=[rectangle,draw=Red]

 \tikzstyle{ma-state}=[rectangle,fill=White,draw=Black]
 \tikzstyle{probability-state}=[circle,fill=Black,draw=Black,inner sep=0pt,minimum size=1pt]

\tikzstyle{arrow-segment-after-probabilities}=[->, >=stealth']
\tikzstyle{arrow-segment-before-probabilities-timed}=[draw=Black,postaction={decorate},
decoration={markings,mark=at position .5 with {\draw[fill=White] (-1pt,-3pt) rectangle (1pt,3pt) ;}}]
\tikzstyle{arrow-segment-before-probabilities-immediate}=[draw=Black,postaction={decorate},
decoration={markings,mark=at position .5 with {\draw[fill=Black] (-1pt,-3pt) rectangle (1pt,3pt) ;}}]

\tikzstyle{ma-trans-immedate}=[->, >=stealth',draw=Black,postaction={decorate},
decoration={markings,mark=at position .5 with {\draw[fill=Black] (-1pt,-3pt) rectangle (1pt,3pt) ;}}]

\tikzstyle{ma-trans-timed}=[->, >=stealth',draw=Black,postaction={decorate},
decoration={markings,mark=at position .5 with {\draw[fill=White] (-1pt,-3pt) rectangle (1pt,3pt) ;}}]

\tikzstyle{gspn-edge}=[->, >=stealth']

\newcommand{\paths}{\mathit{Paths}}
\newcommand{\nnreal}{\mathbb{R}_{\geq 0}}
\newcommand{\F}{\Diamond}
\newcommand{\Act}{\mathit{Act}}

\renewcommand{\it}[1]{\singlearrow{#1}}  
\newcommand{\mt}[1]{\stackrel{#1}{\Longrightarrow}}  

\newcommand{\MS}{\mbox{\sl MS}}
\newcommand{\PS}{\mbox{\sl PS}}
\newcommand{\bfP}{\mathbf{P}}
\newcommand{\bfR}{\mathbf{R}}
\newcommand{\toolname}{{\sc MaMa}}  
\DeclareMathOperator*{\argmin}{arg\,min}
\newcommand{\G}{\Box}
\newcommand{\set}[1]{\{ \, #1 \, \}}

\newcommand{\MaS}{\text{MS}}

\newcommand{\DiBrPr}{\ensuremath{\textrm{\bf P}}} 
\newcommand{\MAM}{\ensuremath{\mathcal{M}}\xspace} 
\newcommand{\dMAM}{\ensuremath{\mathcal{M}_{\delta}}\xspace} 
\newcommand{\GMS}{\ensuremath{\textsl{GM}}\xspace} 
\newcommand{\TAS}{\ensuremath{\mathit{TA}}\xspace} 
\newcommand{\dd}{\:\mathrm{d}}
\newcommand{\ee}{\mathrm{e}}
\newcommand {\defma}{\ensuremath{\mathcal{M}=(S, A, \it{}, \mt{}, s_0)}\xspace}
\newcommand{\rchgls}[1]{\ensuremath{\diamondsuit^{#1}G}\xspace}

\begin{document}

\title[Analysis of Timed and Long-Run Objectives for Markov Automata]{Analysis of Timed and Long-Run Objectives\\for Markov Automata\rsuper*}
    
\author[D.~Guck]{Dennis Guck\rsuper a}	
\address{{\lsuper{a,e}}Formal Methods and Tools, University of Twente, The Netherlands}
\email{\{d.guck,m.timmer\}@utwente.nl}

\author[H.~Hatefi]{Hassan Hatefi\rsuper b}	
\address{{\lsuper{b,c}}Dependable Systems and Software, Saarland University, Germany}	
\email{\{hhatefi,hermanns\}@depend.cs.uni-saarland.de}

\author[H.~Hermanns]{Holger Hermanns\rsuper c}
\address{\vspace{-18 pt}}	

\author[J.-P.~Katoen]{\\Joost-Pieter Katoen\rsuper d}	
\address{{\lsuper d}Formal Methods and Tools, University of Twente, The Netherlands\newline Software Modelling and Verification, RWTH Aachen University, Germany}
\email{katoen@cs.rwth-aachen.de}

\author[M.~Timmer]{Mark Timmer\rsuper e}
\address{\vspace{-18 pt}}


\titlecomment{{\lsuper*}This paper is the extended version of the QEST
  2013 paper entitled ``Modeling, Reduction, and Analysis of Markov
  Automata''~\cite{DBLP:conf/qest/GuckHHKT13}.  The current paper
  focuses on the quantitative analysis of Markov automata, contains
  all full proofs, and has more extensive explanations.}
    
\begin{abstract}
  Markov automata (MAs) extend labelled transition systems with random delays and probabilistic branching.
  Action-labelled transitions are instantaneous and yield a distribution over states, whereas timed transitions 
  impose a random delay governed by an exponential distribution.
  MAs are thus a nondeterministic variation of continuous-time Markov chains.
  MAs are compositional and are used to provide a semantics for engineering frameworks such as (dynamic) fault trees, 
  (generalised) stochastic Petri nets, and the Architecture Analysis \& Design Language (AADL). 
  This paper considers the quantitative analysis of MAs. 
  We consider three objectives: expected time, long-run average, and timed (interval) reachability.  
  Expected time objectives focus on determining the minimal (or maximal) expected time to reach a set of states.
  Long-run objectives determine the fraction of time to be in a set of states when considering
  an infinite time horizon.
  Timed reachability objectives are about computing the probability to reach a set of states within a given time
  interval.
  This paper presents the foundations and details of the algorithms and their correctness proofs.
  We report on several case studies conducted using a prototypical tool implementation of the algorithms, driven by the MAPA modelling language for efficiently generating MAs.
\end{abstract}

\maketitle
\vfill

\section{Introduction} 
Markov automata (MAs, for short) have been introduced in~\cite{EHZ10}
as a continuous-time version of Segala's probabilistic
automata~\cite{Seg95b}. Closed under operators such as parallel
composition and hiding, they provide a compositional formalism for
concurrent soft real time systems. A transition in an MA is either
labelled with a positive real number representing the rate of a
negative exponential distribution, or with an action. An action
transition leads to a discrete probability distribution over states.
MAs can thus model action transitions as in labelled transition
systems, probabilistic branching as found in (discrete time) Markov
chains and Markov decision processes, as well as delays that are
governed by exponential distributions as in continuous-time Markov
chains.

The semantics of MAs has been recently investigated in quite some
detail.  Weak and strong (bi)simulation semantics have been presented
in~\cite{EHZ10,EHZ10b}, whereas it is shown
in~\cite{DBLP:journals/iandc/DengH13} that weak bisimulation provides
a sound and complete proof methodology for reduction barbed
congruence.  A process algebra with data for the efficient modelling
of MAs, accompanied with some reduction techniques using static
analysis, has been presented in~\cite{MAPA}, and model checking of MAs
against Continuous Stochastic Logic (CSL) is discussed
in~\cite{HatefiH12}.  Although the MA model raises several challenging
theoretical issues, both from a semantical and from an analytical
point of view, our main interest is in their practical applicability.
As MAs extend Hermanns' interactive Markov chains
(IMCs)~\cite{Hermanns02}, they inherit IMC application domains,
ranging from GALS hardware designs~\cite{CosteHLS09} and dynamic fault
trees~\cite{DBLP:journals/tdsc/BoudaliCS10} to the standardised
modelling language AADL~\cite{Bozzano,HaverkortKRRS10}.  The
additional feature of probabilistic branching yields additional
expressivity and thereby enriches the spectrum of application contexts
further. This expressivity also makes them a natural semantic model
for other formalisms. Among others, MAs are expressive enough to
provide a natural operational model for generalised stochastic Petri
nets (GSPNs)~\cite{MCB84} and stochastic activity networks
(SANs)~\cite{MeyerMS85}, both popular modelling formalisms for
performance and dependability analysis.  Let us briefly motivate this
by considering GSPNs.  Whereas in SPNs all transitions are subject to
an exponentially distributed delay, GSPNs also incorporate immediate
transitions, transitions that happen instantaneously.
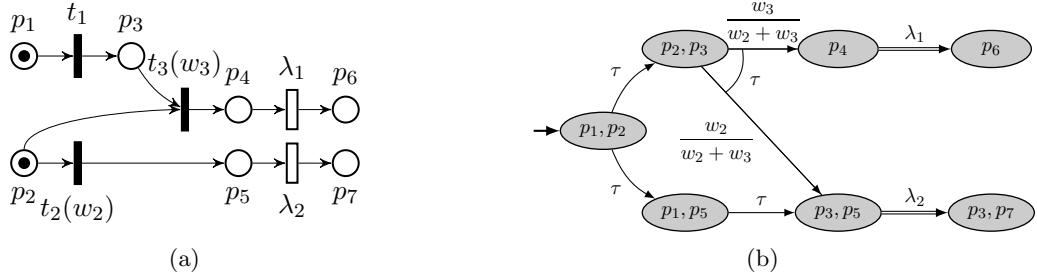
\begin{figure}[t!]
\centering
{}\hfill
\subfigure[]{
\scalebox{0.95}{

\begin{tikzpicture}
	\begin{pgfonlayer}{nodelayer}
		\node [style=place, tokens=1,label=$p_1$] (0) at (0, 0.75) {};
		\node [style=place,label=$p_3$] (1) at (1.5, 0.75) {};
		\node [style=place,label=$p_4$] (2) at (3.0, -0) {};
		\node [style=place,label=below:$p_5$] (3) at (3.0, -0.75) {};
		\node [style=place, tokens=1,label=below:$p_2$] (4) at (0, -0.75) {};
		\node [style=immTH,label=$t_1$] (5) at (0.75, 0.75) {};
		\node [style=immTH,label=$t_3 (w_3)$] (6) at (2.25, -0) {};
		\node [style=immTH,label=below:{$t_2 (w_2)$}] (7) at (0.75, -0.75) {};
		\node [style=stoTH,label=$\lambda_1$] (8) at (3.75, -0) {};
		\node [style=stoTH,label=below:{$\lambda_2$}] (9) at (3.75, -0.75) {};
		\node [style=place,label=$p_6$] (10) at (4.5, -0) {};
		\node [style=place,label=below:$p_7$] (11) at (4.5, -0.75) {};
	\end{pgfonlayer}
	\begin{pgfonlayer}{edgelayer}
		\draw [style=gspn-edge] (0) to (5);
		\draw [style=gspn-edge] (5) to (1);
		\draw [style=gspn-edge, bend right=15, looseness=1.00] (1) to (6);
		\draw [style=gspn-edge, in=180, out=90, looseness=0.50] (4) to (6);
		\draw [style=gspn-edge] (4) to (7);
		\draw [style=gspn-edge] (7) to (3);
		\draw [style=gspn-edge] (6) to (2);
		\draw [style=gspn-edge] (2) to (8);
		\draw [style=gspn-edge] (8) to (10);
		\draw [style=gspn-edge] (3) to (9);
		\draw [style=gspn-edge] (9) to (11);
	\end{pgfonlayer}
\end{tikzpicture}}
\label{fig:gspn-confused}
} \hfill
\subfigure[]{
\tikzstyle{every picture}=[thick, scale=0.78, transform shape]
\tikzstyle{every loop}=[->]
\tikzstyle{every scope}=[>=latex]
\centering
\begin{tikzpicture}[node distance=2.9cm,scale=0.9,every node/.style={transform shape}]
\tikzstyle{every label}=[label distance=0pt]
	\node[initial,state, ellipse, minimum width=43pt,fill=gray!40!white] (p12) {$p_1,p_2$};
	\node[state, ellipse, minimum width=43pt,fill=gray!40!white,above right of=p12,xshift=-0.5cm, yshift=-.5cm] (p23) {$p_2,p_3$};
	\node[state, ellipse, minimum width=43pt,fill=gray!40!white,below right of=p12,xshift=-0.5cm, yshift=.5cm] (p15) {$p_1,p_5$};
	\node[state, ellipse, minimum width=43pt,fill=gray!40!white,right of=p23] (p4) {$p_4$};
	\node[state, ellipse, minimum width=43pt,fill=gray!40!white,right of=p4] (p6) {$p_6$};
	\node[state, ellipse, minimum width=43pt,fill=gray!40!white,right of=p15] (p35) {$p_3,p_5$};
	\node[state, ellipse, minimum width=43pt,fill=gray!40!white,right of=p35] (p37) {$p_3,p_7$};
	
	\path[->] 
		(p4)  edge[thin,double] node[auto] {$\lambda_1$} (p6)
		(p35) edge[thin,double] node[auto] {$\lambda_2$} (p37);
		
	\path[->]
		(p12) edge[thin,bend left=20]  node[auto] {$\tau$} (p23)
		(p12) edge[thin,bend right=20] node[auto,swap] {$\tau$} (p15);
		
	\path[->]
		(p15) edge[thin] node[auto] {$\tau$} (p35);
		
	\path[->]
		(p23) edge[thin] node[auto] {$\dfrac{w_3}{w_2+w_3}$} (p4)
		(p23) edge[thin] node[auto,swap,yshift=0.3cm] {$\dfrac{w_2}{w_2+w_3}$} (p35);
		
	\path[-]
		(p23) edge[thin] node[inner sep=0mm,pos=0.2] (a1) {} (p4)
		(p23) edge[thin] node[inner sep=0mm,pos=0.2] (b1) {} (p35);	
	\path[-,shorten <=-.4pt,shorten >=-.4pt] (a1) edge [thin,bend left]  (b1) node[right,yshift=-.6cm] {$\tau$} ;

\end{tikzpicture}
\hfill{}
\label{fig:gspn-ma-semantics}
}
\caption{(a) Confused GSPN \cite[Fig.\ 21]{Mar95} with partial weights and (b) its MA semantics.}
\end{figure}
The traditional GSPN semantics yields a continuous-time Markov chain
(CTMC), i.e., an MA without action transitions. However, that
semantics is restricted to a subclass of GSPNs, namely those that are
\emph{confusion free}. Confusion~\cite{Mar95} is related to the
presence of nondeterminism. Confused GSPNs are traditionally
considered as semantically ambiguous and thus precluded from any kind
of analysis. This gap is particularly disturbing because several
published semantics for higher-level modelling formalisms---e.g., UML,
AADL, WSDL---map onto GSPN{}s without ensuring the mapping to be free
of confusion, therefore possibly inducing confused models.

It has recently been detailed in~\cite{Katoen12,EHKZ13} that MAs are a
natural semantic model for \emph{every} GSPN.  To give some intuitive
insight into this achievement, consider the GSPN in
Fig.~\ref{fig:gspn-confused}.  This net is confused: In Petri net
jargon, the transitions $t_1$ and $t_2$ are not in conflict, but
firing transition $t_1$ leads to a conflict between $t_2$ and $t_3$,
which does not occur if $t_2$ fires before $t_1$. Though decisive, the
firing order between $t_1$ and $t_2$ is
not determined.  Transitions $t_2$
and $t_3$ are weighted so that in a marking $\{ p_2, p_3 \}$ in which
both transitions are enabled, $t_2$ fires with probability
$\frac{w_2}{w_2{+}w_3}$ and $t_3$ with its complement
probability.  
The weight of transition $t_1$ is not relevant; we assume $t_1$ is not
equipped with a weight.
Classical GSPN semantics and analysis algorithms cannot cope with this
net due to the presence of confusion (i.e., nondeterminism).
Figure~\ref{fig:gspn-ma-semantics} depicts the MA semantics of this
net.  Here, states correspond to sets of net places that contain a
token.  In the initial state, there is a nondeterministic choice
between the transitions $t_1$ and $t_2$.  Note that the presence of
weights is naturally represented by discrete probabilistic branching
as reflected in the outgoing transition from state $\{ p_2, p_3 \}$.
One can show that the MA semantics conservatively extends the
classical semantics, in the sense that the former and the latter are
weakly bisimilar~\cite{EHKZ13} on confusion-free GSPNs.
Thus, if transition $t_1$ in our example is assigned some weight $w_1$, 
the GSPN has no confusion.  This would be reflected in the MA semantics
by replacing the nondeterministic branching in state $\{ p_1, p_2 \}$ by
a single transition, yielding $\{ p_2, p_3 \}$ with probability $\frac{w_1}{w_1{+}w_2}$
and state $\{ p_1, p_5 \}$ with the complement probability.

This paper focuses on the quantitative analysis of MAs---and thus
implicitly of (possibly confused) GSPNs, of AADL specifications containing error models, and
so on.
We present analysis algorithms for three objectives: expected time, long-run average, and timed (interval) reachability.
As the model exhibits nondeterminism, we focus on maximal and minimal values for all three objectives.
We show that expected-time and long-run average objectives can be efficiently reduced to well-known problems on MDPs such as  stochastic shortest path, maximal end-component decomposition, and long-run ratio objectives. 
This generalises (and slightly improves) the results reported in~\cite{DBLP:conf/nfm/GuckHKN12} for IMCs to MAs.
Secondly, we present a discretisation algorithm for timed interval reachability objectives which extends~\cite{DBLP:conf/tacas/ZhangN10}.
Finally, we present the \toolname\ tool chain, an easily accessible publicly available tool chain\footnote{Stand-alone download as well as web-based interface available from \url{http://fmt.cs.utwente.nl/~timmer/mama}.} for the specification, mechanised simplification---such as confluence reduction~\cite{ConfluenceMA}, a form of on-the-fly partial-order reduction---and quantitative evaluation of MAs. 
We describe the overall architectural design, as well as the tool components, and report on empirical results obtained with \toolname\ on a selection of case studies taken from different domains. 
The experiments give insight into the effectiveness of the reduction techniques in \toolname\ and demonstrate that MAs provide the basis of a very expressive stochastic timed modelling approach without sacrificing the ability of time and memory efficient numerical evaluation.

\subsubsection*{Organisation of the paper} 
We introduce Markov automata in Section~\ref{sec:ma}. 
Section~\ref{section:expected} considers the evaluation of expected-time properties. 
Section~\ref{section:longrun} discusses the analysis of long-run properties, and
Section~\ref{section:timed} focuses on timed reachability properties with time-interval bounds. 
Implementation details of our tool, a compositional modelling formalism as well as experimental results are discussed in detail in Section~\ref{sec:tool}. 
Section~\ref{sec:conc} concludes the paper.
We provide the proofs for our main results in the appendix.

\section{Preliminaries}
\label{sec:ma}


%

\subsection{Markov automata}
An MA is a transition system with two types of transitions: probabilistic (as in PAs) and Markovian transitions (as in CTMCs). 
Let $\Act$ be a countable universe of actions with internal action $\tau \in \Act$, and $\distr(S)$ denote the set of discrete probability distribution functions over the countable set $S$.
Let $\alpha$, $\beta$ range over $\Act$ and $\mu, \nu$ over $\distr(S)$.
Actions such as $\alpha$ can be used for interaction with other MAs~\cite{EHZ10}.  
This does not apply to the internal action $\tau$, which is executed autonomously.

\begin{defi}[Markov automaton]\label{def:mas}
A \emph{Markov automaton (MA)} is a tuple $\mathcal{M} = ( S, A, \it{\, },$ $\mt{\, }, s_0)$ where 
$S$ is a nonempty, finite set of states with \emph{initial state} $s_0 \in S$, $A \subseteq \Act$ is a finite set of actions with $\tau \in A$, and 
\begin{itemize}
\item $\it{\, } \subseteq S \times A \times \distr(S)$ is the \emph{probabilistic} transition relation, and
\item $\mt{\, }\ \subseteq S \times \mathbb{R}_{> 0} \times S$ is the \emph{Markovian} transition relation. 
\end{itemize}
\end{defi}

\begin{figure}[t]
\tikzstyle{every picture}=[thick, scale=0.78, transform shape]
\tikzstyle{every loop}=[->]
\tikzstyle{every scope}=[>=latex]
\centering\begin{tikzpicture}[node distance=2.5cm]
\tikzstyle{every label}=[label distance=0pt]

	\node[state, ellipse, minimum size=0pt,fill=gray!40!white] (s000)
{$0,0,0$};
	\node[] (sInit) [left of=s000, node distance=1.5cm] {};
	\draw[->] (sInit) edge [] node [auto, swap] {} (s000);
	\node[state, minimum size=0pt, draw=white] (s100) [above right of=s000]
{};
	\node[state, node distance=0.3cm, ellipse, minimum
size=0pt,fill=gray!40!white] (s100n) [below of=s100] {$1,0,0$};
	\node[state, minimum size=0pt, draw=white] (s010) [below right of=s000]
{};
	\node[state, node distance=0.3cm, ellipse, minimum
size=0pt,fill=gray!40!white] (s010n) [above of=s010] {$0,1,0$};
	\node[state, ellipse, minimum size=0pt,fill=gray!40!white] (s001) [below
right of=s100] {$0,0,1$};
	\node[state, minimum size=0pt, draw=white] (s101) [above right of=s001]
{};
	\node[state, node distance=0.3cm, ellipse, minimum
size=0pt,fill=gray!40!white] (s101n) [below of=s101] {$1,0,1$};
	\node[state, minimum size=0pt, draw=white] (s011) [below right of=s001]
{};
	\node[state, node distance=0.3cm, ellipse, minimum
size=0pt,fill=gray!40!white] (s011n) [above of=s011] {$0,1,1$};
	\node[state, ellipse, minimum size=0pt,fill=gray!40!white] (s111) [below
right of=s101] {$1,1,1$};
	\node[state, ellipse, minimum size=0pt,fill=gray!40!white] (s110) [right
of=s111] {$1,1,0$};

	\draw[->, double,thin] (s000) -- node [auto] {$\lambda_1$}
(s100n);
	\draw[->, double,thin] (s000) --  node [auto, swap]{$\lambda_2$}
(s010n);

	\draw[->,thin] (s100n) edge [] node [auto, swap] {$\frac{9}{10}$} (s001);
	\draw[->,thin] (s100n) edge [bend right=10] node [auto, swap] {$\frac{1}{10}$}
(s101n);
    \draw[thin] (s100n) + (1.25\arclength, -5pt) arc ( -10 :  -39.70265826695821
: 1.25\arclength);
	\path[thin] (s100n) +(31pt, -11.192068269362267pt) node {\phantom{.}} node
{$\tau$} ;

	\draw[->,thin] (s010n) edge [] node [auto] {$\frac{9}{10}$} (s001);
	\draw[->,thin] (s010n) edge [bend left=10] node [auto] {$\frac{1}{10}$}
(s011n);
    \draw[thin] (s010n) + (1.25\arclength, 5pt) arc ( 10 :  39.70265826695821 :
1.25\arclength);
	\path[thin] (s010n) +(31pt, 11.192068269362267pt) node {\phantom{.}} node
{$\tau$} ;

	\draw[->, double,thin] (s001) -- node [auto, swap] {$\mu$} (s000);
	\draw[->, double,thin] (s001) -- node [auto, swap] {$\lambda_1$}
(s101n);
	\draw[->, double,thin] (s001) -- node [auto] {$\lambda_2$}
(s011n);

	\draw[->, double,thin] (s101n) -- node [auto, swap] {$\lambda_2$}
(s111);
	\draw[->, double,thin] (s101n) [bend right=10] to node [auto,
swap] {$\mu$} (s100n);

	\draw[->, double,thin] (s011n) [bend left=10] to node [auto]
{$\mu$} (s010n);
	\draw[->, double,thin] (s011n) -- node [auto] {$\lambda_1$}
(s111);

	\draw[->, double,thin] (s111) -- node [auto] {$\mu$} (s110);

	\draw[->,thin] (s110) edge [in=0, out=120] node [pos=0.65, auto, swap]
{$\frac{9}{10}$} (s101n);
	\draw[->,thin] (s110) edge [in=30, out=120] node [pos=0.65, auto, swap]
{$\frac{1}{10}$} (s111);
    \draw[thin] (s110) + (-20pt,27pt) arc ( 145 :  172 : 0.75\arclength);
	\path[thin] (s110) +(-27pt, 25pt) node {\phantom{.}} node {$\tau$} ;

	\draw[->,thin] (s110) edge [in=0, out=-120] node [pos=0.65, auto]
{$\frac{9}{10}$} (s011n);
	\draw[->,thin] (s110) edge [in=-30, out=-120] node [pos=0.65, auto]
{$\frac{1}{10}$} (s111);
    \draw[thin] (s110) + (-20pt,-27pt) arc ( -145 :  -172 : 0.75\arclength);
	\path[thin] (s110) +(-27pt, -25pt) node {\phantom{.}} node {$\tau$} ;

\end{tikzpicture}
\caption{A queueing system (taken from~\cite{MAPA}), consisting of a server and two stations. 
Each state is represented as a tuple $(s_1,s_2,j)$, with $s_i$ the number of jobs in station $i$, and $j$ the number of jobs
in the server.
The two stations have incoming requests with rates $\lambda_1, \lambda_2$, which are stored until fetched by the server. 
If both stations contain a job, the server chooses nondeterministically (in state (1,1,0)). 
Jobs are processed with rate $\mu$, and when polling a station, with probability $\frac{1}{10}$ the job is erroneously kept in
the station after being fetched. 
For simplicity we assume that each component can hold at most one~job.}
\label{fig:MA}
\end{figure}
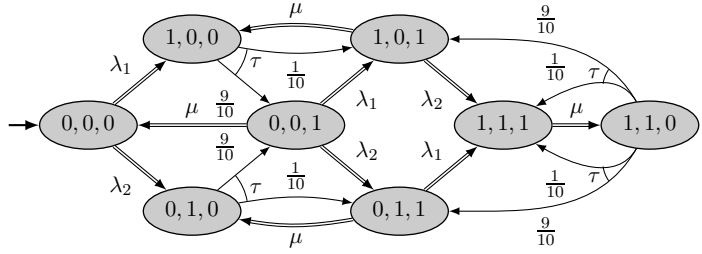

We abbreviate $(s, \alpha, \mu) \in \it{\, }$ by $s \it{\alpha} \mu$ and $(s, \lambda, s') \in \ \mt{\, }$ by $\smash{s \mt{\lambda} s'}$.  
An MA can evolve via its probabilistic and Markovian transitions.
If~$s \it{\alpha} \mu$, it can leave state $s$ by executing the action $\alpha$, after which the probability of going to some state $s' \in S$ is given by $\mu(s')$. 
If $\smash{s \mt{\lambda} s'}$ is the only transition emanating from $s$, a state transition from $s$ to $s'$ can occur after an exponentially distributed delay with rate $\lambda$.
That is to say, the expected delay from $s$ to $s'$ is $\frac{1}{\lambda}$.
If $\smash{s \mt{\lambda} s'}$ and $s \it{\tau} \mu$ for some $\mu$, however, always the $\tau$-transition is taken and never the Markovian one.
This is the \emph{maximal progress} assumption~\cite{EHZ10}.
The rationale behind this assumption is that internal (i.e., $\tau$-labelled) transitions are not subject to interaction and thus can happen immediately, whereas the probability of a Markovian transition to immediately happen is zero. 
Thus, $\smash{s \mt{\lambda} s'}$ almost never fires instantaneously.
Note that the maximal progress assumption does not apply in case $\smash{s \mt{\lambda} s'}$ and $s \it{\alpha} \mu$ with $\alpha \neq \tau$, as $\alpha$-transitions -- unlike $\tau$-transitions -- can be used for synchronisation and thus be subject to a delay.
In this case, the transition $\smash{s \mt{\lambda} s'}$ may happen with positive probability.
The semantics of several Markovian transitions in a state is as follows.
For a state with one or more Markovian transitions, let $\bfR(s,s') = \smash{\sum \{ \lambda \mid s \mt{\lambda} s' \}}$ be the total rate of moving from state~$s$ to state~$s'$, and let $E(s) = \sum_{s' \in S} \ \bfR(s,s')$ be the total outgoing rate of $s$.
If $s$ has more than one outgoing Markovian transition, a \emph{competition} between its Markovian transitions exists.  
Then, the probability of moving from $s$ to state $s'$ within $d$ time units is
\[
\frac{\bfR(s,s')}{E(s)} \cdot \left( 1 - e^{-E(s){\cdot}d} \right).
\]
After a delay of at most $d$ time units (second factor) in state $s$, the MA moves to a direct successor state $s'$ with probability $\bfP(s, s') = \frac{\bfR(s,s')}{E(s)}$. 
Note that also in this case, the maximal progress assumption applies: if $s \it{\tau} \mu$ and $s$ has several Markovian transitions, only the $\tau$-transition can occur and no delay occurs in $s$.
The behaviour of an MA in states with only Markovian transitions is thus the same as in CTMCs~\cite{DBLP:journals/tse/BaierHHK03}.
Fig.~\ref{fig:MA} depicts a sample MA.
Note that this MA only contains $\tau$-labelled probabilistic transitions; by maximal progress, any state has only Markovian transitions or only $\tau$-labelled transitions.
In case several $\tau$-transitions emanate from a state, a nondeterministic choice between these transitions exists.

\subsection{Actions}
Actions different from $\tau$ can be used to compose MAs from smaller MAs using parallel composition.
For instance, $\mathcal{M}_1 \mathbin{||}_H \mathcal{M}_2$ denotes the parallel composition of MA $\mathcal{M}_1$ and $\mathcal{M}_2$ in which actions in the set $H \subseteq \Act$ with $\tau \not\in H$ need to be executed by both MAs simultaneously, and actions not in $H$ are performed autonomously by $\mathcal{M}_i$.
In this paper, we will not cover the details of such composition operation (see~\cite{EHZ10}); it suffices to understand that the distinction between $\tau$ and $\alpha \neq \tau$ is relevant when composing MAs from component MAs.
We assume in the sequel that the MAs to be analysed are \emph{single, monolithic} MAs.
These MAs are not subject to any interaction with other MAs.
Hence, we assume that all transitions are labelled by $\tau$-actions.
(This amounts to the assumption that prior to the analysis all actions needed to compose several MAs are explicitly turned into internal actions by hiding.)
Due to the maximal progress assumption, the outgoing transitions of each state are either all probabilistic or all Markovian.
We can therefore partition the states into a set of probabilistic states, denoted $\PS \subseteq S$, and a set of Markovian states, denoted $\MS \subseteq S$. 
We denote the set of enabled actions in $s$ with $\Act(s)$, where $\Act(s) = \{ \alpha \in A \mid \exists \mu \in \distr(S) \qdot s \it{\alpha}\mu\}$ if $s\in \PS$, and $\Act(s)=\{\bot\}$ otherwise.
 
\subsection{Paths}
A path in an MA is an infinite sequence $ \pi \ = \ s_0 \it{\sigma_0, \mu_0, t_0} s_1 \it{\sigma_1, \mu_1, t_1} \ldots $ with $s_i \in S$,
$\sigma_i = \tau$ or $\sigma_i = \bot$, $\mu_i \in \distr(S)$ and $t_i \in \mathbb{R}_{\geq 0}$.  
For $\sigma_i = \tau$, $s_i \it{\sigma_i, \mu_i, t_i} s_{i+1}$ denotes that after residing $t_i = 0$ time units in $s_i$, the MA moved via action $\sigma_i$ to $s_{i{+}1}$ with probability $\mu_i(s_{i{+}1})$.  
In case $\sigma_i = \bot$, $s_i \it{\bot, \mu_i, t_i} s_{i+1}$ denotes that after residing $t_i$ time units in $s$, a Markovian transition led to $s_{i+1}$ with probability $\mu_i(s_{i+1}) = \bfP(s_i,s_{i+1})$.  
For $t \in \mathbb{R}_{\geq 0}$, let $\pi@t$ denote the \emph{sequence} of states that $\pi$ occupies at time $t$.  
Due to instantaneous probabilistic transitions, $\pi@t$ is a sequence of states, as an MA may occupy various states at the same time instant.  
Let $\paths$ denote the set of infinite paths and $\paths^*$ be the set of finite prefixes thereof (called finite paths).  
The time elapsed along the infinite path $\pi$ is given by $\sum_{i=0}^{\infty}t_i$.  
Path $\pi$ is Zeno whenever this sum converges.  
As the probability of a Zeno path in an MA that only contains Markovian transitions is zero~\cite[Prop.\ 1]{DBLP:journals/tse/BaierHHK03}, an MA is non-Zeno if and only if no SCC with only probabilistic states is reachable (with positive probability).  
As such SCC contains no Markovian transitions, it can be traversed infinitely often without any passage of time. 
In the rest of this paper, we assume MAs to be non-Zeno. 

\subsection{Policies}
Nondeterminism occurs when there is more than one probabilistic transition emanating from a state. 
To define a probability space over sets of infinite paths, we adopt the approach as for MDPs~\cite{Put05} and resolve the nondeterminism by a \emph{policy}.
A policy is a function that yields for each finite path ending in state~$s$ a probability distribution over the set of enabled transitions in $s$. 

Formally, a policy is a function $D \colon \paths^* \to \distr((\Act \mathrel{\cup} \{ \bot \}) \times \distr(S))$. Of course, policies should only choose from available transitions, so we require for each path $\pi$ ending in a state $s_n$ that $D(\pi)(\alpha, \mu) > 0$ implies $s_n \it{\alpha} \mu$ 
and $D(\pi)(\bot, \mu) > 0$ implies that $s_n$ is Markovian and $\mu = \bfP(s_n,\cdot)$.
Let $\GMS$ (generic measurable) denote the most general class of such policies that are still measurable; see~\cite{NSK09} for details on measurability.
In general, a policy randomly picks an enabled action and probability distribution in the final state of a given path.
This is also known as a \emph{history-dependent randomised} policy.
If a policy always selects an action and probability distribution according to a Dirac distribution, it is called a \emph{deterministic} policy.
Policies are also classified based on the level of information they use for the resolution of nondeterminism. 
In the most general setting, a policy may use all information in a finite path, e.g., the states along the path, their ordering in the path, the amount of time spent in each state, and so forth.
A \emph{stationary} policy only bases its decision on the current state, and not on anything else.
That is, $D$ is stationary whenever $D(\pi_1) = D(\pi_2)$ for any finite paths $\pi_1$ and $\pi_2$ that have the same last state. 
A stationary deterministic policy can be viewed as a function $D \colon \PS \rightarrow \Act \times \distr(S)$ that maps each probabilistic state $s$ to an action $\alpha\in\Act$ and probability distribution $\mu\in\distr(S)$ such that $s\it{\alpha}\mu$; such policies always take the same decision every time they are in the same state.
A \emph{time-abstract} policy resolves nondeterminism based on the alternating sequence of states and transitions visited so far, but not on the state residence times.
Let $\TAS$ denote the set of time-abstract policies.
For more details on different classes of policies (and their relationship) on models such as MAs, we refer to~\cite{NSK09}. 
Like for MDPs~\cite{Put05}, a stationary or time-abstract policy on an MA induces a countable stochastic process that is equivalent to a (continuous-time) Markov chain.
Using a standard cylinder-set construction on infinite paths in such Markov chains~\cite{DBLP:journals/tse/BaierHHK03} we obtain a $\sigma$-algebra of subsets of $\paths$; given a policy~$D$ and an initial  state~$s$, a measurable set of paths is equipped with probability measure $\Pr_{s,D}$.

To ease the development of the theory, and without loss of generality, we assume that each internal action induces a unique probability distribution. Note that this is no restriction: if there are multiple $\tau$-transitions emerging from a state $s\in \PS$, we may replace the $\tau$ by internal actions $\tau_1$ to $\tau_{n}$, where $n$ is the out-degree of $s$.

\subsection{Stochastic shortest path (SSP) problems}
As some objectives on MAs can be reduced to SSP problems, we briefly introduce them.
An MDP is a tuple $(S,A,\bfP,s_0)$ where $S$ is a finite set of states, $A \subseteq \Act$ is a set of actions, $\bfP \colon S \times A \times S \to [0,1]$ such that for each state $s$ and each $\alpha$, $\sum_{s' \in S} \bfP(s,\alpha,s') \in \set{0,1}$, and $s_0 \in S$ is the initial state. 
It is assumed that in each state at least one action is enabled, i.e., $\bfP(s,\alpha,s') > 0$ for each $s$, for some~$\alpha$.
A \emph{non-negative SSP problem} is a tuple $(S,A,\bfP,s_0, G,c,g)$ where the first four elements represent its underlying MDP accompanied by a set $G \subseteq S$ of goal states, cost function $c \colon (S \setminus G) \times A \to \mathbb{R}_{\geq 0}$ and terminal cost function $g\colon G \to \mathbb{R}_{\geq 0}$. 
A path through an MDP is an alternating sequence $s_0 \it{\alpha_0} s_1 \it{\alpha_1} \ldots$ such that $\bfP(s_i, \alpha_i, s_{i{+}1}) > 0$, for all $i$.
The accumulated cost along a path~$\pi$ through the MDP before reaching $G$, denoted by $C_G(\pi)$, is $\sum_{j{=}0}^{k{-}1} c(s_j,\alpha_j) + g(s_k)$ where $k$ is the state index of reaching $G$.
If $\pi$ does not reach $G$, then $C_G(\pi)$ equals $\infty$.
As standard in MDPs~\cite{Put05}, nondeterminism between different actions in a state is resolved using policies; similar to the notion for MAs, a stationary deterministic policy is a function $D\colon\PS\to\Act$.
Let $\mathit{cR}^{\min}(s, \diamondsuit G)$ denote the minimum expected cost reachability of $G$ in the SSP (under all policies) when starting from $s$.
It is a well-known result that stationary policies suffice to achieve $\mathit{cR}^{\min}(s, \diamondsuit G)$.
This expected cost can be obtained by solving an LP (linear programming) problem~\cite{BerTsi91}.

\section{Expected time objectives}
\label{section:expected}

Let $\mathcal{M}$ be an MA with state space $S$ and $G \subseteq S$ a set of goal states.
Define the (extended) random variable $V_G \colon \paths \rightarrow \nnreal^{\infty}$ as the elapsed time before first visiting some state in $G$. That is, for an infinite path $\pi = s_0 \smash{\xrightarrow{\sigma_0,\mu_0,t_0}} s_1 \smash{\xrightarrow{\sigma_1,\mu_1,t_1}} \cdots$, let $V_G(\pi) = \min \left\{ t \in \nnreal \mid G \cap \pi@t \not= \emptyset \right\}$ where $\min (\emptyset) = \infty$.  
(With slight abuse of notation we use $\pi@t$ as the set of states occurring in the sequence $\pi@t$.)
The minimal expected time to reach $G$ from $s \in S$ is defined by
\begin{align*}
  \mathit{eT}^{\min}(s, \diamondsuit G) \ = \ 
  \inf_{D\in \GMS} \mathbb{E}_{s,D}(V_G) \ = \ 
  \inf_{D\in \GMS} \int_{\paths} \hspace{-2ex} V_G(\pi) \cdot \Pr\nolimits_{s,D}(\mathrm{d}\pi)
\end{align*}
where $D$ is a generic measurable policy on $\mathcal{M}$.
(In the sequel, we assume that $\mathit{eT}^{\min}$ is a function indexed by $G$.)
Note that by definition of $V_G$, only the amount of time before entering the first $G$-state is relevant.
Hence, we may turn all $G$-states into absorbing without affecting the expected time reachability.
It is done via replacing all of their emanating transitions by a single Markovian self loop (a Markovian transition to the state itself) with an arbitrary rate. 
In the remainder of this section we assume all goal states to be absorbing.
Let $\mu^s_\alpha$ be the distribution such that $s \it{\alpha} \mu^s_\alpha$.
As we assume that all action labels of the transitions emanating a state are unique (by numbering them), this distribution is unique.

\begin{restatable}{thm}{thmExpectedReachability}\label{thm_expected_reachability}
  The function $\mathit{eT}^{\min}$ is a fixpoint of the Bellman operator
  {\small \begin{align*}
    \left[L(v)\right](s) = \begin{cases}
      \displaystyle \frac{1}{E(s)} + \sum_{s' \in S} \bfP(s,s') \cdot v(s') & \text{ if } s \in \MS \setminus G \\
      \displaystyle \min_{\alpha \in \textit{\footnotesize Act}(s)} \sum_{s' \in S} \mu^s_\alpha(s') \cdot v(s') & \text{ if } s \in \PS \setminus G \\
      \displaystyle 0 & \text{ if } s \in G,
    \end{cases}
  \end{align*}}%
  where $\Act(s)=\{\tau_i \mid s\it{\tau_i}\mu\}$ and $\mu_\alpha^s \in \distr(S)$ is as formerly defined.
\end{restatable}
We will later see that $\mathit{eT}^{\min}$ is in fact the unique fixpoint of the Bellman operator.
Let us explain the above result.
For a goal state, the expected time obviously is zero.
For a Markovian state $s \not\in G$, the minimal expected time to reach some state in $G$ is the expected sojourn time in $s$ (which equals $\frac{1}{E(s)}$) plus the expected time to reach some state in $G$ via one of its successor states.
For a probabilistic state, an action is selected that minimises the expected time according to the distribution $\mu^s_\alpha$ corresponding to $\alpha$ in state $s$.
The characterisation of $\mathit{eT}^{\min}(s, \diamondsuit G)$ in Thm.\,\ref{thm_expected_reachability} allows us to
reduce the problem of computing the minimum expected time reachability in an MA \mbox{to a non-negative 
SSP problem~\cite{BerTsi91,deAlf99}.}
This goes as follows.

\begin{defi}[SSP for minimum expected time reachability]\label{def_expected_time_reachability_ssp}
The SSP of MA $\mathcal{M} = \left( S, A, \it{\, }, \mt{\, }, s_0 \right)$ for the expected time reachability of $G \subseteq S$ is
\[
     {\sf ssp}_{et}(\mathcal{M}) = \left( S, A \cup \left\{ \bot \right\}, \bfP, s_0, G,c, g \right)
\]
where $g(s) = 0$ for all $s \in G$ and
  {\small \begin{align*}
    \bfP(s, \sigma, s') & = \begin{cases}
      \frac{\bfR(s,s')}{E(s)} & \text{if } s \in \MS, \sigma = \bot \\
      \mu^s_\sigma(s') & \text{if } s \in \PS, s \it{\sigma} \mu^s_\sigma \\
      0 & \text{otherwise, and}
    \end{cases} &
    c(s,\sigma) & = \begin{cases}
      \frac{1}{E(s)} & \text{if } s \in \MS \setminus G, \sigma = \bot \\
      0 & \text{otherwise.}
    \end{cases}
  \end{align*}}
\end{defi}
Terminal costs are zero.
Transition probabilities are defined in the standard way.
The cost of a Markovian state is its expected sojourn time, whereas that of a probabilistic one is zero.
\begin{restatable}{thm}{thmExpectedTimeReachabilityReduction}\label{thm_expected_time_reachability_reduction}
Given an MA $\mathcal{M}$, $\mathit{eT}^{\min}(s, \diamondsuit G)$ equals $\mathit{cR}^{\min}(s, \diamondsuit G)$ in 
$\mbox{\sf ssp}_{et}(\mathcal{M})$.
\end{restatable}
Thus there is a stationary deterministic policy on $\mathcal{M}$ yielding $\mathit{eT}^{\min}(s, \diamondsuit G)$.
Moreover, the uniqueness of the minimum expected cost of an SSP~\cite{BerTsi91,deAlf99} now yields that
$\mathit{eT}^{\min}(s, \diamondsuit G)$ is the unique fixpoint of $L$ (see Thm.~\ref{thm_expected_reachability}). This follows from the fact that the Bellman operator defined in Thm~\ref{thm_expected_reachability} equals the Bellman operator for $\mathit{cR}^{\min}(s, \diamondsuit G)$.
The uniqueness result enables the usage of standard solution techniques such as value iteration and linear programming to compute $\mathit{eT}^{\min}(s, \diamondsuit G)$.
For maximum expected time objectives, a similar fixpoint theorem is obtained, and it can be proven that those objectives correspond to the maximal expected reward in the SSP problem defined above.
Thus far, we have assumed MAs to be non-Zeno, i.e., they do not contain a reachable cycle solely consisting of probabilistic transitions.
However, the above notions can all be extended to deal with such Zeno cycles, by, e.g., setting the minimal expected time of states in Zeno BSCCs that do not contain $G$-states to be infinite (as such states cannot reach $G$).  
Similarly, the maximal expected time of states in Zeno end components (that do not contain $G$-states) can be defined as infinity, as in the worst case these states will~never~reach~$G$.

\newcommand{\mI}{\mathcal{M}}
\newcommand{\bdI}{\mathbf{1}}
\newcommand{\LRA}{\textit{LRA}}
\newcommand{\mR}{\mathcal{R}}
\newcommand{\Paths}{\paths}

\section{Long-run objectives}
\label{section:longrun}
Let $\mI$ be an MA with state space $S$ and $G \subseteq S$ a set of goal states.
Let $\bdI_G$ be the characteristic function of $G$ on finite sequences, i.e., $\bdI_G(\pi) = 1$ if and only if $s \in G$ for some $s$ in $\pi$.
Following the ideas of \cite{DBLP:conf/lics/Alfaro98,LHK01}, the fraction of time spent in $G$ on an infinite path $\pi$ in $\mI$ up to time bound $t \in \mathbb{R}_{\geq 0}$ is given by the random variable
$A_{G,t}(\pi) \ = \ \frac{1}{t} \int_0^t \bdI_G(\pi@u)\,\mathrm{d}u$.
Taking the limit $t \rightarrow \infty$, we obtain the random variable
\begin{align*}
A_{G}(\pi) \ = \ \lim_{t\to\infty} A_{G,t}(\pi) \ = \ \lim_{t\to\infty} \frac{1}{t}\int_0^t \bdI_G(\pi@u)\,\mathrm{d}u.
\end{align*}
The expectation of $A_{G}$ for policy $D$ and initial state $s$ yields the corresponding long-run average time spent in $G$:
\begin{align*}
\LRA^D(s, G) 
= 
\mathbb{E}_{s,D}(A_{G}) 
= 
\int_{\paths} \hspace{-2ex} A_{G}(\pi) \cdot {\Pr}_{s,D}(\mathrm{d}\pi).
\end{align*}
The minimum long-run average time spent in $G$ starting from state $s$ is then:
\begin{align*}
\LRA^{\min}(s,G) \ = \ \inf_{D\in \GMS}\ \LRA^D(s, G) \ = \ \inf_{D\in \GMS} \mathbb{E}_{s,D}(A_{G}).
\end{align*}
Note that $\bdI_G(\pi@u) = 1$ if and only if $\pi@u$ is a sequence containing at least one state in $G$. For the long-run average analysis, we assume w.l.o.g.\ that $G\subseteq \MS$, as the long-run average time spent in any probabilistic state is always 0. 
This claim follows directly from the fact that probabilistic states are instantaneous, i.e.\ their sojourn time is $0$ by definition.
Note that in contrast to the expected time analysis, $G$-states cannot be made absorbing in the long-run average analysis.

First we need to introduce \emph{maximal end components}. A sub-MA of MA $\mathcal{M}$ is a pair $(S',K)$ where $S' \subseteq S$ and $K\colon S' \to 2^A$ is a function such that: (i) $K(s) \neq \emptyset$, (ii) $s \in S'$ and $\alpha \in K(s)$ and $s \it{\alpha} \mu$ with $\mu(s') > 0$ implies $s' \in S'$, and (iii) $s \in S'$ and $\smash{s \mt{\lambda} s'}$ implies $s' \in S'$.
A sub-MA $(S',K)$ is contained in a sub-MA $(S'',K')$ if $S' \subseteq S''$ and $K(s) \subseteq K'(s)$ for all $s \in S'$.
An \emph{end component} is a sub-MA whose underlying graph is strongly connected; it is \emph{maximal} w.r.t.\ $K$ if it is not contained in any other end component $(S'',K')$ of $\mathcal{M}$.

In the remainder of this section, we discuss in detail how to compute the minimum long-run average fraction of time spent in $G$ in an MA $\mI$ with initial state $s_0$.
The general idea is the following three-step procedure:
\begin{enumerate}
\item Determine the maximal end components $\{ \mI_1, \ldots, \mI_k \}$ of MA $\mI$.
\item Determine $\LRA^{\min}(G)$ in maximal end component $\mI_j$ for all $j \in \{ 1, \ldots, k \}$.  
\item Reduce the computation of $\LRA^{\min}(s_0,G)$ in MA $\mI$ to an SSP problem.
\end{enumerate}
The first phase can be performed by a graph-based algorithm~\cite{deA97_thesis,DBLP:conf/soda/ChatterjeeH11}, whereas the last two phases boil down to solving (distinct) LP problems.

\subsection{Unichain MA}
We first show that for unichain MAs 
computing $\LRA^{\min}(s, G)$ can be reduced to determining long-run ratio objectives in MDPs. 
The notion of unichain is standard in MDPs~\cite{Put05} and is adopted to MAs in a straightforward manner.
An MA is unichain if for any stationary deterministic policy the induced stochastic process consists of a single ergodic class plus a possibly non-empty set of transient states\footnote{State $s$ is \emph{transient} if and only if the probability of the set of paths that start from $s$ but never return back to it is positive, otherwise it is \emph{recurrent}. An MA is \emph{ergodic} if for all stationary deterministic policies the induced stochastic process consists of a single recurrent class.}.
Let us first explain the long-run ratio objectives. Let $M=(S,A,\bfP,s_0)$ be an MDP.
Assume w.l.o.g.\ that for each $s\in S$ there exists $\alpha\in A$ such that $\bfP(s,\alpha,s') > 0$ for some $s'\in S$.
Let $c_1, c_2\colon S \times A \to \mathbb{R}_{\geq 0}$ be cost functions.
The operational interpretation is that a cost $c_1(s,\alpha)$ is incurred when selecting action $\alpha$ in state $s$, and similar for $c_2$.
Our interest is the \emph{ratio} between $c_1$ and $c_2$ along a path.
The \emph{long-run ratio} $\mathcal{R}$ between the accumulated costs $c_1$ and $c_2$ along the infinite path $\pi = s_0 \it{\alpha_0} s_1 \it{\alpha_1} \ldots$ in the MDP $M$ is defined by:  
$$
\mathcal{R}(\pi) \ = \ \displaystyle \lim_{n \to \infty} \dfrac{\sum_{i=0}^{n-1} c_1(s_i,\alpha_i)}{\sum_{j=0}^{n-1} c_2(s_j,\alpha_j)}.
$$ 
The minimum long-run ratio objective for state $s$ of MDP $M$ is defined by:
\begin{align*}
  R^{\min}(s) \ = \ 
  \inf_D \mathbb{E}_{s,D}(\mR) \ = \ 
  \inf_D \sum_{\pi \in \Paths} \mR(\pi) \cdot \text{Pr}_{s,D}(\pi).
\end{align*}

Here, $\Paths$ is the set of paths in the MDP, $D$ is a stationary deterministic MDP-policy, and $\Pr$ is the probability measure on MDP-paths.
From~\cite[Th.\ 6.14]{deA97_thesis}, it follows that $R^{\min}(s)$ can be obtained by solving the following LP problem with real variables $k$ and non-negative $x_s$ for each $s \in S$: Maximise $k$ subject to:
$$
x_s \, \leq \, c_1(s,\alpha) - k \cdot c_2(s,\alpha) + \sum_{s' \in S} \bfP(s,\alpha,s') \cdot x_{s'} \quad
\mbox{ for each } s \in S, \alpha \in A.
$$
We now transform an MA into an MDP with two cost functions as follows.
\begin{defi}[From MA to 2-cost MDPs]\label{def:MAtomdp}
Let $\mI = \left( S, A,  \it{\, }, \mt{\, }, s_0 \right)$ be an MA and $G \subseteq S$ a set of goal states.
The MDP $\mbox{\sf mdp}(\mI) = (S, A\cup\{\bot\}, \bfP, s_0)$, where $\bfP$ is defined as in Def.~\ref{def_expected_time_reachability_ssp}, 
is extended with cost functions $c_1$ and $c_2$ defined by:
{\small \begin{align*}
  c_1(s,\sigma) & = \begin{cases}
    \frac{1}{E(s)} & \text{if } s \in \MS \cap G \wedge \sigma = \bot \\
    0 & \text{otherwise,}
  \end{cases}
  &
  c_2(s,\sigma) & = \begin{cases}
    \frac{1}{E(s)} & \text{if } s \in \MS \wedge \sigma = \bot \\
    0 & \text{otherwise.}
  \end{cases}
\end{align*}}
\end{defi}
Observe that cost function $c_2$ keeps track of the average residence
time in state~$s$ whereas $c_1$ only does so for states in $G$.
Furthermore, $\mathcal{R}$ is well-defined in this setting, since the
cost functions $c_1$ and $c_2$ are obtained from non-Zeno MA.  In
other words, the probability of the set of paths with
ill-defined long-run ratio is zero.

\begin{restatable}{thm}{unichainTheorem}\label{thm_unichain_theorem}
For unichain MA $\mI$, $LRA^{\min}(s,G)$ equals $R^{\min}(s)$ in $\mbox{\sf mdp}(\mI)$.
\end{restatable}

To summarise, computing the minimum long-run average fraction of time that is spent in some goal state in $G \subseteq S$ in a unichain MA $\mI$ equals the minimum long-run ratio objective in an MDP with two cost functions.
The latter can be obtained by solving an LP problem.
Observe that for any two states $s$, $s'$ in a unichain MA, $\LRA^{\min}(s,G)$ and $\LRA^{\min}(s',G)$ coincide.
We therefore omit the state and simply write $\LRA^{\min}(G)$ when considering unichain MAs.

\subsection{Arbitrary MA}
Let $\mI$ be an MA with initial state $s_0$ and maximal end components $\{ \mI_1, \ldots,$ $\mI_k \}$ for $k > 0$ where MA $\mI_j$ has state space $S_j$. 
\newtheorem{lemma}[thm]{Lemma}
\begin{restatable}{lemma}{lemMEC}\label{lem:lemMEC}
Let $\mI$ be a maximal end component and $D$ a stationary deterministic policy inducing a multichain on $\mI$. Then there exists a stationary deterministic policy $D'$ inducing a unichain on $\mI$ such that the long-run ratio is at least as good as for $D$.
\end{restatable}
Therefore, we can say that each $\mI_j$ induces a unichain MA for the optimal long-run ratio.
Using this decomposition of $\mI$ into maximal end components, we obtain the following result:

\newif\ifnote\notetrue
\begin{restatable}{thm}{lemLRASSP}\ifnote\label{lem:LRA_SSP}\footnote{This theorem corrects a small flaw in the corresponding theorem for IMCs in~\cite{DBLP:conf/nfm/GuckHKN12}.}\fi
For MA $\mI = (S, A, \it{\,}, \mt{\, }, s_0)$ with MECs $\{ \mI_1, \ldots, \mI_k \}$ 
with state spaces $S_1, \dots, S_k \subseteq S$, and set of goal states $G \subseteq S$:
\begin{align*}
  \LRA^{\min}(s_0,G) & = \inf_{D\in GM}\sum_{j=1}^{k} \LRA^{\min}_j(G) \cdot {\Pr}_{s_0,D}(\diamondsuit \Box S_j),
\end{align*}
where ${\Pr}_{s_0,D}(\diamondsuit \Box S_j)$ is the probability to eventually reach and continuously stay in some states in $S_j$ from $s_0$ under policy $D$ and $\LRA^{\min}_j(G)$ is the LRA of $G \cap S_j$ in unichain~MA~$\mI_j$.  
\end{restatable}
\ifnote\notefalse

Computing the minimal LRA for arbitrary MAs is now reducible to a non-negative SSP problem.
This proceeds as follows.
In MA $\mI$, we replace each maximal end component $\mathcal{M}_j$ by two fresh states $q_j$ and $u_j$.
Intuitively, $q_j$ represents $\mathcal{M}_j$ whereas $u_j$ can be seen as the gate to and from $\mathcal{M}_j$.
Thus, state $u_j$ has a Dirac transition to $q_j$ as well as all probabilistic transitions leaving $S_j$.
Let $U$ denote the set of $u_j$ states and $Q$ the set of $q_j$ states.
For simplicity of the definition we assume w.l.o.g. that each probabilistic state induces a $\tau$-transition with an index of the state. Further, the $\tau$-transitions of each state $s_k\in \PS$ are numbered from $1$ to $n_{s_k}\in\mathbb{N}$, where $n_{s_k}$ is the number of probability distributions induced by $\tau_{s_k}$. Thus, we denote an action in state $s_k$ with $\tau_{{s_k}_l}$ with $l\in \{1\ldots n_{s_k}\}$.
\begin{defi}[SSP for long-run average]\label{defSSP}
The SSP of MA $\mathcal{M}$ for the LRA in $G \subseteq S$ is $\mbox{\sf ssp}_{lra}(\mathcal{M}) = \left( (S \setminus \smash{\bigcup_{i=1}^k} S_i) \cup U \cup Q, A \cup \{ \bot \}, \bfP', s_0, Q, c, g \right)$, where $g(q_i) = \LRA^{\min}_i(G)$ for $q_i \in Q$ and $c(s,\sigma) = 0$ for all $s$ and $\sigma\in A \cup\{\bot\}$. $\bfP'$ is defined as follows. Let $S' =  S \setminus \smash{\bigcup_{i=1}^k} S_i$.  $\bfP'(s,\sigma,s')$ equals $\bfP(s,\sigma,s')$ for all $s,s' \in S'$ and $\sigma\in A\cup\{\bot\}$.
For the new states $u_j$:
{\small
  \begin{align*}
  \bfP'(u_j, \tau_{{s_k}_l}, s') & =  \bfP(s_k, \tau_{{s_k}_l}, s') \quad \text{\!if } s' \in S' \wedge s_k \in S_j\wedge l\in\{1\dotso n_{s_k}\} 
  &
  \mbox{\!\!and\!\! }\\
  \quad \bfP'(u_i, \tau_{{s_k}_l}, u_j) & = \bfP(s_k, \tau_{{s_k}_l}, S_j) \quad \text{\!if } s_k \in S_i\wedge l\in\{1\dotso n_{s_k}\}\wedge \tau_{{s_k}_i}\not\in A_i
\end{align*}}%
Finally, we have: $\bfP'(q_j,\bot,q_j) = 1 = \bfP'(u_j, \bot,q_j)$ and $\bfP'(s, \sigma, u_j) = \bfP(s, \sigma, S_j)$.
\end{defi}
Here, $\bfP(s,\alpha, S')$ is a shorthand for $\sum_{s' \in S'} \bfP(s,\alpha,s')$ and $A_i$ denotes the action set of maximal end component $\mathcal{M}_i$.
The terminal costs of the new $q_i$-states are set to $\LRA^{\min}_i(G)$.

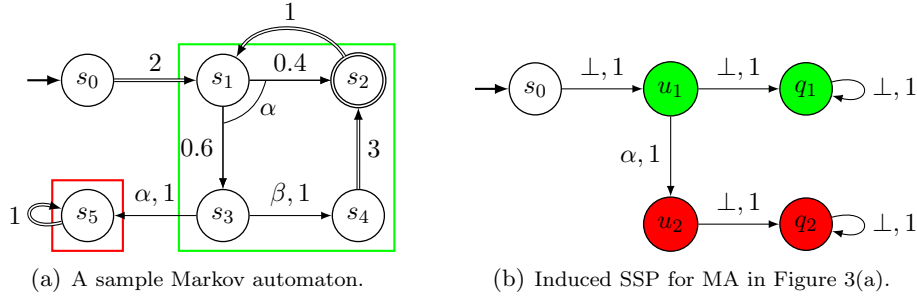
\begin{figure}[t!]
\centering
\subfigure[\scriptsize{A sample Markov automaton.}]{\label{fig:example_mec}
\begin{tikzpicture}[scale=0.9, every node/.style={transform shape}]
	\node[state, initial] (s0) {$s_0$};
	\node[state] (s3) [right of=s0,node distance=2cm] {$s_1$};
	\node[state] (s5) [below of=s3,node distance=2cm] {$s_3$};
	\node[state, accepting] (s6) [right of=s3,node distance=2cm] {$s_2$};
	\node[state] (s8) [left of=s5,node distance=2cm] {$s_5$};
	\node[state] (s9) [right of=s5,node distance=2cm] {$s_4$};
	\node[rectangle] (tmp) [left of=s3, node distance=.38cm] {}; 
	\node[rectangle] (tmp2) [left of=s5, node distance=.38cm] {}; 
	
	\node[rectangle,draw,green,fit=(tmp)(tmp2)(s6)(s9)] (mec1) {};
	\node[rectangle,draw,red,fit=(s8)] (mec2) {};
	
	\path[->] (s0) edge[double,thin] node[sloped,above] {$2$} (s3);
	\path[->] (s3) edge[thin] node[left] {$0.6$} (s5);
	\path[->] (s3) edge[thin] node[sloped,above] {$0.4$} (s6);
	
	\path[->]
		(s3) edge[thin] node[inner sep=0mm,pos=0.2] (a1) {} (s5)
		(s3) edge[thin] node[inner sep=0mm,pos=0.2] (b1) {} (s6);	
	\path[-,shorten <=-.4pt,shorten >=-.4pt] (a1) edge [thin,bend right]  (b1) node[right,yshift=.2cm,xshift=.4cm] {$\alpha$} ;
	
	\path[->] (s5) edge[thin] node[sloped,above] {$\alpha,1$} (s8);
	\path[->] (s5) edge[thin] node[sloped, above] {$\beta,1$} (s9);
	\path[->] (s8) edge[thin,double,loop left] node[left] {$1$} (s8);
	\path[->] (s9) edge[thin,double] node[right] {$3$} (s6);
	\path[->] (s6) edge[thin,double, bend right=60] node[sloped,above] {$1$} (s3);
\end{tikzpicture}
}
\hspace{.3cm}
\subfigure[\scriptsize{Induced SSP for MA in Figure~\ref{fig:example_mec}.}]{\label{fig:example_ssp}
\begin{tikzpicture}[scale=0.9, every node/.style={transform shape}]
	\node[state, initial] (s0) {$s_0$};
	\node[state,fill=green] (u2) [right of=s0,node distance=2cm] {$u_1$};
	\node[state,fill=green] (q2) [right of=u2,node distance=2cm] {$q_1$};
	\node[state,fill=red] (u1) [below of=u2,node distance=2cm] {$u_2$};
	\node[state,fill=red] (q1) [right of=u1,node distance=2cm] {$q_2$};
	
	\path[->] (s0) edge[thin] node[sloped,above] {$\bot,1$} (u2);
	\path[->] (u2) edge[thin] node[sloped,above] {$\bot,1$} (q2);
	\path[->] (u2) edge[thin] node[left] {$\alpha,1$} (u1);
	\path[->] (u1) edge[thin] node[sloped,above] {$\bot,1$} (q1);
	\path[->] (q2) edge[thin,loop right] node[right] {$\bot,1$} (q2);
	\path[->] (q1) edge[thin,loop right] node[right] {$\bot,1$} (q1);
\end{tikzpicture}
}
\caption {Example for Definition~\ref{defSSP}.}
\end{figure}
\begin{exa}
Consider the MA $\mI$ from Figure~\ref{fig:example_mec}, having MECs $\mI_1$ with $S_1=\{s_1,s_2,s_3,s_4\}$ and $\mI_2$ with $S_2=\{s_5\}$. For the simplification of the action notation, we use $\alpha$ and $\beta$ instead of $\tau$. Let $G = \{ s_2 \}$.  By Definition~\ref{defSSP}, $\mbox{\sf ssp}_{lra}(\mI)$ is defined as follows.  As $k{=}2$, $U=\{u_1,u_2\}$ and $Q=\{q_1,q_2\}$. Hence, $S_{\mbox{\sf ssp}} = \{s_0,u_1,u_2,q_1,q_2\}$. First consider $s,s'\in S'$. Since, $S'=\{s_0\}$ and there exists no transition from $s_0$ to $s_0$ we can omit the first rule. Now consider all outgoing transitions from MECs. For $\mI_1$ there exists a transition from $s_3 \it{\alpha,1} s_5$ in the underlying MA, where $s_3\in S_1$ and $s_5 \in S_2$. It follows that $\bfP'(u_1,\alpha,u_2)=\bfP(s_3,\alpha,S_2)=1$. 
Now consider all states in $U$ and $Q$ and add new transitions with $\bfP(u_i,\bot,q_i) = \bfP(q_i,\bot,q_i)=1$ for $i=1,2$. Finally, consider all states $s \in S_{\mbox{\sf ssp}} \cap S$ with a transition into a MEC. Hence, $\bfP'(s_0,\bot,u_1)=\bfP(s_0,\bot,s_1)=1$. The MDP of $\mbox{\sf ssp}_{lra}(\mI)$ is depicted in Figure~\ref{fig:example_ssp}.
 \label{ex:append}
\end{exa}

\begin{restatable}{thm}{thmLRASSP}\label{thm:LRA_SSP}
For MA $\mathcal{M}$, $\LRA^{\min}(s_0,G)$ equals $cR^{\min}(s_0 ,\diamondsuit Q)$ in SSP $\mbox{\sf ssp}_{lra}(\mathcal{M})$.
\end{restatable}
To summarise, computing the minimum long-run average fraction of time that is spent in some goal states in $G \subseteq S$ in an arbitrary MA $\mI$ starting in state $s_0$ equals the minimum expected cost of an SSP.

\section{Timed reachability objectives}
\label{section:timed}


  This section presents an algorithm that approximates time-bounded reachability probabilities
  in MAs.  We start with a fixpoint characterisation, and then explain how 
  these probabilities can be approximated using a discretisation technique.

\subsection{Fixpoint characterisation}
Our goal is to come up with a fixpoint characterisation for the maximum (or minimum) probability to reach a set of goal states in a time interval. Let $\mathcal{I}$ and $\mathcal{Q}$ be the set of all nonempty nonnegative real intervals with real and rational bounds, respectively.
For interval $I \in \mathcal{I}$ and $t \in \mathbb{R}_{\ge 0}$, let $I \ominus t = \left \{ x - t \mid x \in I \wedge x \ge t \right \}$. 
Given MA $\mathcal{M}$, $I \in \mathcal{I}$ and a set $G \subseteq S$ of goal states, the set of all paths that reach some goal states within interval $I$ is denoted by $\diamondsuit^{I} \, G$. 
Let $p^{\mathcal{M}}_{\max}(s,\diamondsuit^I \, G)$ be the maximum probability of reaching $G$ within interval $I$ if starting in state $s$ at time $0$. 
Here, the maximum is taken over all possible generic measurable policies.
The next lemma provides a characterisation of $p_{\max}^{\mathcal{M}}(s,\diamondsuit^I \, G)$ as a fixpoint.
\begin{lem} 
\label{fpc:ma} 
Let $\mathcal{M}$ be an MA, $G \subseteq S$ and $I \in \mathcal{I}$ with $\inf I=a$ and $\sup I=b$. Then, $p^{\mathcal{M}}_{\max}(s,\diamondsuit^I \, G)$ is the least fixpoint of the higher-order operator
    $\Omega\colon (S \times \mathcal{I} \rightarrow [0,1])
    \rightarrow (S \times \mathcal{I} \rightarrow [0,1])$,
    which for $s \in \MS$ is given by:
    \begin{align*}
      \Omega(F)(s,I)&=
        \begin{cases}
          \displaystyle\int_{0}^{b}E(s)\ee^{-E(s)t}\sum_{s' \in S}\bfP(s,s')F(s',I \ominus t)\dd t & \!s \notin G \\
          \displaystyle\ee^{-E(s)a} + \int_{0}^{a}E(s)\ee^{-E(s)t}\sum_{s' \in
            S}\bfP(s,s')F(s',I \ominus t)\dd t & \!s \in G
        \end{cases}
\intertext{and for $s \in \PS$ is defined by:}
        \Omega(F)(s,I)&=
        \begin{cases}
          1 & s \in G \wedge  0\in I \\
          \max_{\alpha\in\Act(s)} \sum_{s' \in S}
          \mu^s_\alpha(s')F(s',I) & \mathrm{otherwise.}
        \end{cases}
      \end{align*} 
  \end{lem}
  The proof of Lemma~\ref{fpc:ma} is a slight adaptation
  of the proof of~\cite[Thm.~4]{Fu13}, where it has been also shown
  that $p^{\mathcal{M}}_{\max}(s,\diamondsuit^I \, G)$ is Lipschitz
  continuous and thus measurable. The characterisation is a simple
  generalisation of that for IMCs~\cite{DBLP:conf/tacas/ZhangN10},
  reflecting the fact that taking an action from a probabilistic state
  leads to a distribution over the states (rather than a single
  state).  The above characterisation yields a Volterra integral
  equation system which is in general not directly tractable
  \cite{DBLP:journals/tse/BaierHHK03}.  To tackle this problem, we
  approximate the fixpoint characterisation using discretisation,
  extending ideas developed in \cite{DBLP:conf/tacas/ZhangN10}. 
  
  \subsection{Discretisation}
  We
  split the time interval into equally-sized discretisation steps,
  each of length $\delta$.  The discretisation step is assumed to be
  small enough such that with high probability it carries at most one
  Markovian transition.  This allows us to construct a discretised MA
  (dMA), a variant of a semi-MDP, obtained by summarising the
  behaviour of the MA at equidistant time points.  Paths in a dMA can
  be seen as time-abstract paths in the corresponding MA, implicitly
  counting discretisation steps, and thus discrete time.
\begin{defi}
  Given MA \defma and discretisation step
  $\delta\in\mathbb{R}_{>0}$, $\mathcal{M}_{\delta}=( S, A, \it{\,
  },$ $\mt{\, }_{\delta}, s_0)$ is the dMA induced from $\mathcal{M}$
  with respect to $\delta$, with $\mt{}_{\delta} \, = \{ \, (s,
  \mu^{s})\mid s \in \MS \, \}$, where
    \begin{equation*}
      \mu^s (s') = 
      \begin{cases}
        (1-\ee^{-E(s)\delta})\bfP(s,s') & \mbox{if } s' \neq s\\
        (1-\ee^{-E(s)\delta})\bfP(s,s') + \ee^{-E(s)\delta} & \mbox{otherwise.}
      \end{cases}
    \end{equation*}
\end{defi} 
Using the above fixpoint characterisation, it is now possible to
relate reachability probabilities in the MA $\mathcal{M}$ to
reachability probabilities in its dMA $\mathcal{M}_{\delta}$.
\begin{restatable}{thm}{thmTBR}\label{thm:tbr}
    Given MA $\mathcal{M}=(S,A, \it{}, \mt{}, s_0)$, $G \subseteq S$, interval 
    $I=[0,b] \in \mathcal{Q}$ with $b \ge 0$ and $\lambda = \max_{s \in \scriptsize\MS}E(s)$. 
    Let $\delta\in\mathbb{R}_{>0}$ be such that $b=k_b\delta$ for some $k_b \in \mathbb{N}$. 
    Then, for all $s \in S$ it holds that
    \begin{equation*}
      p^{\mathcal{M}_{\delta}}_{\max}(s, \diamondsuit^{[0,k_b]} \, G) 
      \ \leq \ 
      p^{\mathcal{M}}_{\max}(s, \diamondsuit^{[0,b]} \, G) 
      \ \leq \ 
      p^{\mathcal{M}_{\delta}}_{\max}(s, \diamondsuit^{[0,k_b]} \, G) + 1 - e^{- \lambda b}\big(1+ \lambda \delta\big)^{k_b}.
    \end{equation*}
  \end{restatable}
  This theorem can be extended to intervals with non-zero lower
  bounds; for the sake of brevity, the details are omitted here.
%
  The remaining problem is to compute
  $p^{\dMAM}_{\max}(s,\diamondsuit^{[0,k_b]}\,G)$, which is the
  maximum probability to reach some goal state in dMA \dMAM within the
  step bound $k_b$ from initial state $s$.  Let
  $\diamondsuit^{[0,k_b]}\,G$ be the set of infinite (time-abstract)
  paths of \dMAM that reach some state in $G$ within $k_b$ steps; the
  objective is then formalised by
  $p^{\dMAM}_{\max}(s,\diamondsuit^{[0,k_b]}\,G)=\sup_{D \in \TAS}
  {\Pr}_{s,D}(\diamondsuit^{[0,k_b]}\,G)$ where we recall that $\TAS$ denotes the
  set of time-abstract policies.  Our algorithm
  is now an adaptation (to dMA) of the well-known value iteration
  \mbox{scheme for MDPs.}

The algorithm proceeds by backward unfolding of the dMA in an
iterative manner, starting from the goal states.  Each iteration
intertwines the analysis of Markovian states and of probabilistic states.
The key idea is that a path from probabilistic states to $G$ is split
into two parts: reaching Markovian states from probabilistic states in zero
time and reaching goal states from Markovian states in interval $[0,j]$,
where $j$ is the step count of the iteration. The former computation
can be reduced to an unbounded reachability problem in the MDP induced
by probabilistic states with rewards on Markovian states.  For the latter,
the algorithm operates on the previously computed reachability
probabilities from all Markovian states up to step count $j$.  We can
generalise this recipe from step-bounded reachability to step
interval-bounded reachability; details are described
in~\cite{HatefiH12}.

\section{Tool chain and case studies}
\label{sec:tool}

This section describes the implementation of the algorithms discussed, together with the modelling features resulting in our \toolname~tool chain. 
Also, we present two case studies that provide empirical evidence of the strengths and weaknesses \mbox{of the \toolname~tool chain.}

\subsection{Modelling}\label{sec:modelling}
As argued in the introduction, MAs can be used as a semantical model for various modelling formalisms.
We use the process-algebraic specification language MAPA (Markov Automata Process Algebra)~\cite{MAPA,markPhd}.
This language contains the usual process algebra operators, can treat data as first-class citizens, and supports several reduction techniques for MA specifications.
In fact, it turns out to be beneficial to map a language (like GSPNs) to MAPA so as to profit from these reductions.

The MAPA language supports algebraic processes featuring data, nondeterministic choice, action prefix with probabilistic choice, rate prefix, conditional behaviour and process instantiation (allowing recursion). Using MAPA processes as basic building blocks, the language also supports the modular construction of large systems via top-level parallelism, encapsulation, hiding and renaming. 
The operational semantics of a MAPA specification yields an MA; for a detailed exposition of the syntax and semantics we refer to~\cite{MAPA,markPhd}.

To enable state space reduction and generation, our tool chain uses a linearised normal form of MAPA referred to as MLPE (Markovian Linear Probabilistic process Equation).
In this format, there is precisely one process which consists of a nondeterministic choice between a set of symbolic transitions, making MLPEs easy to translate to MAs.
Every MAPA specification can be translated efficiently into an MLPE while preserving strong bisimulation~\cite{MAPA}.

\subsubsection*{Reduction techniques}
On MLPEs, several reduction techniques have been defined. Some of them simplify the MLPE to improve readability and speed up state space generation, while others really modify it in such a way that the underlying MA gets smaller. Being defined on the specification, these reductions eliminate the need to ever generate the original unreduced state space. We briefly discuss six such techniques.

\begin{itemize}
\item 
\emph{Maximal progress reduction} removes Markovian transitions from states also having $\tau$-transitions (motivated by the maximal progress assumption). 
\item
\emph{Constant elimination}~\cite{KPST11} replaces parameters that remain forever constant by their initial (and hence permanent) value.
\item 
\emph{Expression simplification}~\cite{KPST11} evaluates functions for which all parameters are constants and applies basic laws from logic. 
\item 
\emph{Summation elimination}~\cite{KPST11} removes trivial nondeterministic choices often arising from synchronisations.
\item
\emph{Dead-variable reduction}~\cite{PT09} detects parts of the specification in which the value of some variable is irrelevant: it will be overwritten before being used for all possible futures. When reaching such a part, the variable is reset to its initial value.
\item
\emph{Confluence reduction}~\cite{ConfluenceMA} detects spurious nondeterminism resulting from parallel composition. It denotes a subset of the probabilistic transitions of a MAPA specification as confluent, meaning that they can safely be given priority if enabled together with other transitions. 
\end{itemize}

\subsection{\toolname~tool chain} 

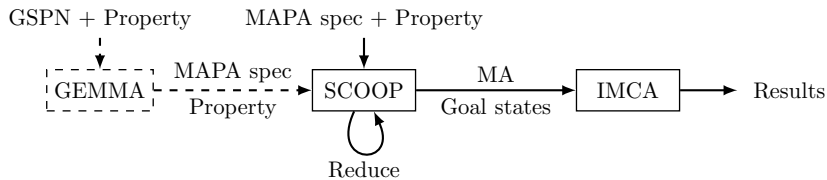
\begin{figure}[b]
\centering\begin{tikzpicture}[scale=0.78, transform shape]

	\node[state, rectangle, minimum width=50pt, minimum height=20pt] (s_0) {SCOOP};
	\node[state, rectangle, minimum width=50pt, minimum height=20pt] (s_2) [right of=s_0, node distance=4.5cm] {IMCA};
	\node[state, rectangle, draw=white, minimum width=50pt, minimum height=20pt] (s_3) [right of=s_2, node distance=2.75cm] {Results};
	\node[state, rectangle, draw=white] (s_4) [above of=s_0, node distance=1.2cm] {MAPA spec + Property};
	\draw[draw=white] (s_0) -- node [auto,swap] {Goal states} (s_2);
	\draw[->] (s_0) -- node [auto] {MA} (s_2);
	\draw[->, in=295, out=245, loop] (s_0) edge node [auto,swap] {Reduce} (s_0);
	\draw[->] (s_2) -- node [auto] {} (s_3);
	\draw[->] (s_4) -- (s_0);

	\node[state, dashed, rectangle, minimum width=50pt, minimum height=20pt] (s_5) [left of=s_0, node distance=4.5cm] {GEMMA};
	\draw[draw=white] (s_5) -- node [auto,swap] {Property} (s_0);
	\draw[->, dashed] (s_5) -- node [auto] {MAPA spec} (s_0);
	\node[state, rectangle, draw=white] (s_6) [above of=s_5, node distance=1.2cm] {GSPN + Property};
	\draw[->, dashed] (s_6) -- (s_5);

\end{tikzpicture}
\caption{Analysing Markov automata using the \toolname~tool chain.}
\label{fig:approach}
\end{figure}

Our tool chain consists of several tool components: SCOOP~\cite{Timmer11,MAPA}, IMCA~\cite{DBLP:conf/nfm/GuckHKN12}, and GEMMA~\cite{gemma}, see Figure~\ref{fig:approach}.
The tool chain comprises about 8,000 LOC (without comments).
SCOOP (written in Haskell) supports the generation of MAs from MAPA specifications by a translation into the MLPE format.
It implements all the reduction techniques described above.
The capabilities of the IMCA tool component (written in {\tt C++}) have been lifted to expected time and long-run objectives for MAs, and extended with timed reachability objectives.
It also supports (untimed) reachability objectives which are not treated further here.
A prototypical translator from GSPNs to MAs, in fact MAPA specifications, has been realised (the GEMMA component, written in Haskell).
We connected the three components into a single tool chain, by making SCOOP export the (reduced) state space of an MLPE in the IMCA input language. 
Additionally, SCOOP has been extended to translate properties, based on the actions and parameters of a MAPA specification, to a set of goal states in the underlying MA. 
That way, in one easy process, systems and their properties can be modelled in MAPA, translated to an optimised MLPE by SCOOP, exported to the IMCA tool and then analysed.

\subsection{Case studies}

This section presents experiments with \toolname. All experiments were conducted on a 2.5 GHz Intel Core i5 processor with 4GB RAM, running Mac OS X~10.8.3. 

\subsubsection*{Processor grid} 
First, we consider a model of a $2 \times 2$ concurrent processor architecture. 
Using GEMMA \cite{gemma}, we automatically derived the MA model from the GSPN model in~\cite[Fig.~11.7]{Mar95}. 
Previous analysis of this model required weights for all immediate transitions, which necessitates having complete knowledge of the mutual behaviour of all these transitions. 
We allow a weight assignment to just a (possibly empty) subset of the immediate transitions---reflecting the practical scenario of only knowing the mutual behaviour for a selection of the transitions. 
For this case study we indeed kept weights for only a few of the transitions, obtaining probabilistic behaviour for them and nondeterministic behaviour for the others.

Table~\ref{tab:grid_tb} reports on the time-bounded and time-interval bounded probabilities for reaching a state such that the first processor has an empty task queue. We vary the degree of multitasking $K$, the error bound $\epsilon$ and the interval~$I$. For each setting, we report the number of states $|S|$ and goal states $|G|$, and the 
generation time with SCOOP (both with and without the reductions from Section~\ref{sec:modelling}).

The runtime demands grow with both the upper and lower time
bound, as well as with the required accuracy. The model size also
affects the per-iteration cost and thus the overall complexity of
reachability computation.
Note that the reductions speed-up the analysis times by a factor between $1.8$ and $2.5$: even more than the reduction in state space size. This is due to the fact that these techniques significantly reduce the degree of nondeterminism.

Table~\ref{tab:grid} displays the results for expected time until an empty task queue, as well as the long-run average that a processor is active. 
In contrast to~\cite{Mar95}, which fixes all nondeterminism and obtains, for instance, an LRA of $0.903$ for $K=2$, we are now able to retain nondeterminism and provide the more informative interval $[0.8810, 0.9953]$.
Again, SCOOP's reduction techniques significantly improve runtimes.

  \begin{table}[t]
\centering
{\vskip15pt
\hspace{-0.5cm}\scalebox{.75}{\begin{tabular}{c||ccc|ccc|cc|rrr|rrr}
& \multicolumn{3}{c|}{unreduced} & \multicolumn{3}{c|}{reduced}\\
$K$ & $|S|$ & $|G|$ & time & $|S|$ & $|G|$ & time & $\epsilon$ & $I$ &  \smash{\begin{turn}{35}$p^{\min}(s_0, \diamondsuit^{I} G)$ \end{turn}}\hspace*{-10ex}& \smash{\begin{turn}{35}time(unred)\end{turn}}\hspace*{-8ex} & \smash{\begin{turn}{35}time(red)\end{turn}}\hspace*{-6ex} &\smash{\begin{turn}{35} $p^{\max}(s_0, \diamondsuit^{I} G)$ \end{turn}}\hspace*{-10.5ex}&  \smash{\begin{turn}{35}time(unred)\end{turn}} \hspace*{-9ex} &  \smash{\begin{turn}{35}time(red)\end{turn}} \hspace*{-7ex}\\
\hline
\hline
\multirow{4}{*}{2} & \multirow{4}{*}{2{,}508}  & \multirow{4}{*}{1{,}398} & \multirow{4}{*}{0.6} & 
\multirow{4}{*}{1{,}789}  & \multirow{4}{*}{1{,}122} & \multirow{4}{*}{0.8} & 
$10^{-2}$ & $[0,3]$ & 0.91\phantom{0} & 58.5 & 31.0 & 0.95\phantom{0} & 54.9 & 21.7\\
  & & & & & & & 
$10^{-2}$ & $[0,4]$ & 0.96\phantom{0} & 103.0 & 54.7 & 0.98\phantom{0} & 97.3 & 38.8\\
  & & & & & & & 
$10^{-2}$ & $[1,4]$ & 0.91\phantom{0} & 117.3 & 64.4 & 0.96\phantom{0} & 109.9 & 49.0\\
  & & & & & & & 
$10^{-3}$ & $[0,3]$ & 0.910 & 580.1 & 309.4 & 0.950 & 544.3 & 218.4\\
\hline
\multirow{4}{*}{3} & \multirow{4}{*}{10{,}852}  & \multirow{4}{*}{4{,}504} & \multirow{4}{*}{3.1} & 
\multirow{4}{*}{7{,}201}  & \multirow{4}{*}{3{,}613} & \multirow{4}{*}{3.5} & 
$10^{-2}$ & $[0,3]$ & 0.18\phantom{6} & 361.5 & 202.8 & 0.23\phantom{1} & 382.8 & 161.1\\
  & & & & & & & 
$10^{-2}$ & $[0,4]$ & 0.23\phantom{6} & 643.1 & 360.0 & 0.30\phantom{1} & 681.4 & 286.0\\
  & & & & & & & 
$10^{-2}$ & $[1,4]$ & 0.18\phantom{6} & 666.6 & 377.3 & 0.25\phantom{1} & 696.4 & 317.7\\
  & & & & & & & 
$10^{-3}$ & $[0,3]$ & 0.176 & 3{,}619.5 & 2{,}032.1 & 0.231 & 3{,}837.3 & 1{,}611.9\\
\hline
4 & 31{,}832  & 10{,}424 & 9.8 & 20{,}021  & 8{,}357 & 10.5 & 
$10^{-2}$ & $[0,3]$ & 0.01\phantom{6} & 1{,}156.8 & 614.9 & 0.03\phantom{1} & 1{,}196.5 & 486.4\\
\hline
\end{tabular}}
}
\caption{Interval reachability probabilities for the grid. (Time in seconds.)}
\label{tab:grid_tb}
\end{table}

  \begin{table}[t]
\centering
{
\vskip10mm\hspace{-0.3cm}\scalebox{.8}{\begin{tabular}{c||rrr|rrr|rrr|rrr}
$K$ & \smash{\begin{turn}{35}$eT^{\min}(s_0, \F G)$ \end{turn}}\hspace*{-10ex}& \smash{\begin{turn}{35}time(unred)\end{turn}}\hspace*{-8ex} & \smash{\begin{turn}{35}time(red)\end{turn}}\hspace*{-6ex} &\smash{\begin{turn}{35} $eT^{\max}(s_0, \F G)$ \end{turn}}\hspace*{-10.5ex}&  \smash{\begin{turn}{35}time(unred)\end{turn}} \hspace*{-9ex} &  \smash{\begin{turn}{35}time(red)\end{turn}} \hspace*{-7ex} & 
\smash{\begin{turn}{35}$\LRA^{\min}(s_0, G)$ \end{turn}}\hspace*{-10ex}& \smash{\begin{turn}{35}time(unred)\end{turn}}\hspace*{-8ex} & \smash{\begin{turn}{35}time(red)\end{turn}}\hspace*{-6ex} &\smash{\begin{turn}{35} $\LRA^{\max}(s_0, G)$ \end{turn}}\hspace*{-10.5ex}&  \smash{\begin{turn}{35}time(unred)\end{turn}} \hspace*{-9ex} &  \smash{\begin{turn}{35}time(red)\end{turn}} \hspace*{-7ex}
\\
\hline
\hline
2 & 1.0000 & 0.3 & 0.1 & 1.2330 & 0.7 & 0.3 & 0.8110 & 1.3 & 0.7 & 0.9953 & 0.5 & 0.2\\
3 & 11.1168 & 18.3 & 7.7 & 15.2768 & 135.4 & 40.6 & 0.8173 & 36.1 & 16.1 & 0.9998 & 4.7 & 2.6\\ 
4 & 102.1921 & 527.1 & 209.9 & 287.8616 & 6{,}695.2 & 1{,}869.7 & 0.8181 & 505.1 & 222.3 & 1.0000 & 57.0 & 34.5\\ 
\hline
\end{tabular}}
}
\caption{Expected times and long-run averages for the grid. (Time in seconds.)}
\label{tab:grid}
\end{table}

\begin{table}[t]
\centering
{
\vskip15pt\hspace{-0.4cm}\scalebox{.75}{\begin{tabular}{cc||ccc|ccc|cc|rrr|rrr}
&& \multicolumn{3}{c|}{unreduced} & \multicolumn{3}{c|}{reduced}\\
$Q$ & $N$ & $|S|$ & $|G|$ & time & $|S|$ & $|G|$ & time & $\epsilon$ & $I$ &  \smash{\begin{turn}{35}$p^{\min}(s_0, \diamondsuit^{I} G)$ \end{turn}}\hspace*{-10ex}& \smash{\begin{turn}{35}time(unred)\end{turn}}\hspace*{-8ex} & \smash{\begin{turn}{35}time(red)\end{turn}}\hspace*{-6ex} &\smash{\begin{turn}{35} $p^{\max}(s_0, \diamondsuit^{I} G)$ \end{turn}}\hspace*{-10.5ex}&  \smash{\begin{turn}{35}time(unred)\end{turn}} \hspace*{-9ex} &  \smash{\begin{turn}{35}time(red)\end{turn}} \hspace*{-7ex}\\
\hline
\hline
\multirow{2}{*}{2} & \multirow{2}{*}{3} & \multirow{2}{*}{1{,}497} & \multirow{2}{*}{567} & \multirow{2}{*}{0.4} & \multirow{2}{*}{990} & \multirow{2}{*}{324} & \multirow{2}{*}{0.2} &  
$10^{-3}$ & $[0,1]$ & 0.277 & 4.7 & 2.9 & 0.558 & 4.6 & 2.5\\
& & & & & & & &
$10^{-3}$ & $[1,2]$ & 0.486 & 22.1 & 14.9 & 0.917 & 22.7 & 12.5\\
\hline
\multirow{2}{*}{2} & \multirow{2}{*}{4} & \multirow{2}{*}{4{,}811} & \multirow{2}{*}{2{,}304} & \multirow{2}{*}{1.0} & \multirow{2}{*}{3{,}047} & \multirow{2}{*}{1{,}280} & \multirow{2}{*}{0.6} &  
$10^{-3}$ & $[0,1]$ & 0.201 & 25.1 & 14.4 & 0.558 & 24.0 & 13.5 \\
& & & & & & & &
$10^{-3}$ & $[1,2]$ & 0.344 & 106.1 & 65.8 & 0.917 & 102.5 & 60.5\\
\hline
\hline
\multirow{2}{*}{3} & \multirow{2}{*}{3} & \multirow{2}{*}{14{,}322} & \multirow{2}{*}{5{,}103} & \multirow{2}{*}{3.0} & \multirow{2}{*}{9{,}522} & \multirow{2}{*}{2{,}916} & \multirow{2}{*}{1.7} &  
$10^{-3}$ & $[0,1]$ & 0.090 & 66.2 & 40.4 & 0.291 & 60.0 & 38.5\\
& & & & & & & &
$10^{-3}$ & $[1,2]$ & 0.249 & 248.1 & 180.9 & 0.811 & 241.9 & 158.8\\
\hline
\multirow{2}{*}{3} & \multirow{2}{*}{4} & \multirow{2}{*}{79{,}307} & \multirow{2}{*}{36{,}864} & \multirow{2}{*}{51.6} & \multirow{2}{*}{50{,}407} & \multirow{2}{*}{20{,}480} & \multirow{2}{*}{19.1} &  
$10^{-3}$ & $[0,1]$ & 0.054 & 541.6 & 303.6 & 0.291 & 578.2 & 311.0\\
& & & & & & & &
$10^{-3}$ & $[1,2]$ & 0.141 & 2{,}289.3 & 1{,}305.0 & 0.811 & 2{,}201.5 & 1{,}225.9\\
\hline
\hline
\multirow{2}{*}{4} & \multirow{2}{*}{2} & \multirow{2}{*}{6{,}667} & \multirow{2}{*}{1{,}280} & \multirow{2}{*}{1.1} & \multirow{2}{*}{4{,}745} & \multirow{2}{*}{768} & \multirow{2}{*}{0.8} &  
$10^{-3}$ & $[0,1]$ & 0.049 & 19.6 & 14.0 & 0.118 & 19.7 & 12.8\\
& & & & & & & &
$10^{-3}$ & $[1,2]$ & 0.240 & 83.2 & 58.7 & 0.651 & 80.9 & 53.1\\
\hline
\multirow{2}{*}{4} & \multirow{2}{*}{3} & \multirow{2}{*}{131{,}529} & \multirow{2}{*}{45{,}927} & \multirow{2}{*}{85.2} & \multirow{2}{*}{87{,}606} & \multirow{2}{*}{26{,}244} & \multirow{2}{*}{30.8} &  
$10^{-3}$ & $[0,1]$ & 0.025 & 835.3 & 479.0 & 0.118 & 800.7 & 466.1\\
& & & & & & & &
$10^{-3}$ & $[1,2]$ & 0.114 & 3{,}535.5 & 2{,}062.3 & 0.651 & 3{,}358.9 & 2{,}099.5\\
\hline
\end{tabular}
}
}
\caption{Interval reachability probabilities for the polling system. (Time in seconds.)}
\label{tab:poll_job_tb}
\end{table}


  \begin{table}[t]
\centering
{\vskip20pt
\hspace{-0.3cm}\scalebox{.75}{\begin{tabular}{cc||rrr|rrr|rrr|rrr}
$Q$ & $N$ & \smash{\begin{turn}{35}$eT^{\min}(s_0, \F G)$ \end{turn}}\hspace*{-10ex}& \smash{\begin{turn}{35}time(unred)\end{turn}}\hspace*{-8ex} & \smash{\begin{turn}{35}time(red)\end{turn}}\hspace*{-6ex} &\smash{\begin{turn}{35} $eT^{\max}(s_0, \F G)$ \end{turn}}\hspace*{-10.5ex}&  \smash{\begin{turn}{35}time(unred)\end{turn}} \hspace*{-9ex} &  \smash{\begin{turn}{35}time(red)\end{turn}} \hspace*{-7ex} & 
\smash{\begin{turn}{35}$\LRA^{\min}(s_0, G)$ \end{turn}}\hspace*{-10ex}& \smash{\begin{turn}{35}time(unred)\end{turn}}\hspace*{-8ex} & \smash{\begin{turn}{35}time(red)\end{turn}}\hspace*{-6ex} &\smash{\begin{turn}{35} $\LRA^{\max}(s_0, G)$ \end{turn}}\hspace*{-10.5ex}&  \smash{\begin{turn}{35}time(unred)\end{turn}} \hspace*{-9ex} &  \smash{\begin{turn}{35}time(red)\end{turn}} \hspace*{-7ex}
\\
\hline
\hline
2 & 3 & 1.0478 & 0.2 & 0.1 & 2.2489 & 0.3 & 0.2 & 0.1230 & 0.8 & 0.5 & 0.6596 & 0.2 & 0.1\\
2 & 4 & 1.0478 & 0.2 & 0.1 & 3.2053 & 2.0 & 1.0 & 0.0635 & 9.0 & 5.2 & 0.6596 & 1.3 & 0.6\\
\hline
3 & 3 & 1.4425 & 1.0 & 0.6 & 4.6685 & 8.4 & 5.0 & 0.0689 & 177.9 & 123.6 & 0.6600 & 26.2 & 13.0 \\
3 & 4 & 1.4425 & 9.7 & 4.6 & 8.0294 & 117.4 & 67.2 & 0.0277 & 7{,}696.7 & 5{,}959.5 & 0.6600 & 1{,}537.2 & 862.4\\
\hline
4 & 2 & 1.8226 & 0.4 & 0.3 & 4.6032 & 2.4 & 1.6 & 0.1312 & 45.6 & 32.5 & 0.6601 & 5.6 & 3.9\\
4 & 3 & 1.8226 & 29.8 & 14.2 & 9.0300 & 232.8 & 130.8 & \multicolumn{3}{c|}{-- timeout (18 hours) --} & 0.6601 & 5{,}339.8 & 3{,}099.0\\
\hline
\end{tabular}}
}
\caption{Expected times and long-run averages for the polling system. (Time in seconds.)}
\label{tab:poll_job}
\end{table}

\subsubsection*{Polling system} 
Second, we consider a polling system with two stations and one server,
similar to the one depicted in Figure~\ref{fig:MA} and inspired
by~\cite{Polling}.  There are incoming requests of $N$ possible types,
each of them with a (possibly different) service rate. Additionally,
the stations each store these in a local queue of size $Q$.  We vary the values
of~$Q$ and~$N$, analysing a total of six different settings.
Since---as for the previous case---analysis scales proportionally with
the error bound, we \mbox{keep this constant here}.

Table~\ref{tab:poll_job_tb} reports results for
time-bounded and time-interval bounded properties, and
Table~\ref{tab:poll_job} displays probabilities and runtime
results for expected times and long-run averages. For all analyses,
the goal set consists of all states for which both station queues are full.

\section{Conclusion}
\label{sec:conc}

This paper presented new algorithms for the quantitative analysis of Markov automata (MAs) and proved their correctness.
Three objectives have been considered: expected time, long-run average, and timed reachability.
The \toolname\ tool chain supports the modelling and reduction of MAs, and can analyse these three objectives.
It is also equipped with a prototypical tool to map GSPNs onto MAs.
The \toolname\ tool is accessible via its easy-to-use web interface that can be found at \mbox{\url{http://fmt.cs.utwente.nl/~timmer/mama}}.
Experimental results on a processor grid and a polling system give insight into the accuracy and scalability of the presented algorithms.
Future work will focus on efficiency improvements and reward extensions~\cite{atvapaper}.

\section*{Acknowledgements}
This work is funded by the EU FP7-projects SENSATION and MEALS, the STW project ArRangeer (grant 12238), the NWO project SYRUP (grant 612.063.817), and the DFG Sonderforschungsbereich AVACS.

\bibliographystyle{abbrv}
\bibliography{references}

\appendix

\section{Proof of Theorem~\ref{thm_expected_reachability}}

Recall that the minimal expected time to reach $G$ from $s \in S$ is defined by
\begin{align*}
  \mathit{eT}^{\min}(s, \diamondsuit G) \ = \
  \inf_{D\in \GMS} \mathbb{E}_{s,D}(V_G) \ = \
  \inf_{D\in \GMS} \int_{\paths} \hspace{-2ex} V_G(\pi) \; \Pr\nolimits_{s,D}(\mathrm{d}\pi)
\end{align*}
where $D$ is a generic measurable policy on $\mathcal{M}$.  $\mathit{eT}^{\min}$ is a function indexed by $G$. Further, $V_G \colon \paths \rightarrow \nnreal^{\infty}$ is the elapsed time before visiting some state in $G$ for the first time, i.e., $V_G(\pi) = \min \left\{ t \in \nnreal \mid G \cap \pi@t \not= \emptyset \right\}$ where $\min (\emptyset) = \infty$. Let $\Delta(\pi,k)=\sum_{i=0}^{k-1}t_i$ be the elapsed time on infinite path $\pi=s_0\it{\sigma_0, \mu_0, t_0}s_1\it{\sigma_1,\mu_1,t_1}\ldots$ after $k$ steps. $V_G$ can be therefore rewritten as
\begin{equation}
V_G(\pi)=\begin{cases}
\Delta(\pi,j) & \text{ if } \pi[j] \in G\wedge \forall i<j.\ \pi[i]\not\in G\\
\Delta(\pi,\infty) & \text{ if } \forall i.\ \pi[i]\not\in G
\end{cases}
\end{equation}
\thmExpectedReachability*
\proof We show that $L(eT^{\min}(s,\diamondsuit G))=eT^{\min}(s,\diamondsuit G)$, for all $s\in S$. Therefore, we will distinguish three cases: $s\in\MS\setminus G, s\in\PS\setminus G$, and $s\in G$. Note that $D\in \GMS$.
\renewcommand{\labelenumi}{(\roman{enumi})}
\begin{enumerate}
	\item if $s\in\MS\setminus G$, we derive
	{\footnotesize
		\begin{eqnarray*}
			eT^{\min}(s,\diamondsuit G) & = & \inf_D \mathbb{E}_{s,D}(V_G) = \inf_D \int_{\paths} V_G(\pi)\cdot {\Pr}_{s,D} (\mathrm{d}\pi) = \inf_D \int_{\paths} \Delta(\pi,k)\cdot{\Pr}_{s,D} (\mathrm{d}\pi)\\
			& = & \inf_D\int_{\paths}\sum_{i=0}^{k-1}t_i \cdot {\Pr}_{s,D}(\mathrm{d}\pi) = \inf_D\int_{\paths}(t_0+\sum_{i=1}^{k-1}t_i) \cdot {\Pr}_{s,D}(\mathrm{d}\pi)\\
			& = & \inf_D \int_{0}^{\infty} t \cdot E(s)\ee^{-E(s)t} + \sum_{s' \in S} \bfP(s,s') \cdot \mathbb{E}_{s',D[s\it{\bot,\bfP(s,\cdot),t} s']} (V_G)\dd t\\
			& = & \inf_D \left(\int_{0}^{\infty} t \cdot E(s)\ee^{-E(s)t}\dd t + \int_{0}^{\infty}\sum_{s' \in S} \bfP(s,s') \cdot \mathbb{E}_{s',D[s\it{\bot,\bfP(s,\cdot),t} s']} (V_G)\dd t\right)\\
			& = & \inf_D \left(\frac{1}{E(s)} + \sum_{s' \in S}\bfP(s,s') \cdot \int_{0}^{\infty} \mathbb{E}_{s',D[s\it{\bot,\bfP(s,\cdot),t} s']} (V_G)\dd t\right)\\
			& = &  \frac{1}{E(s)} + \inf_D \sum_{s' \in S}\bfP(s,s') \cdot \int_{0}^{\infty} \mathbb{E}_{s',D[s\it{\bot,\bfP(s,\cdot),t} s']} (V_G)\dd t\\
			& = & \frac{1}{E(s)} + \inf_D \sum_{s' \in S} \bfP(s,s') \cdot \mathbb{E}_{s',D} (V_G)\\
			& = & \frac{1}{E(s)} + \sum_{s' \in S} \bfP(s,s') \cdot \inf_D \mathbb{E}_{s',D} (V_G)\\
			& = & \frac{1}{E(s)} + \sum_{s' \in S} \bfP(s,s') \cdot eT^{\min}(s',\diamondsuit G)\\
			& = & L(eT^{\min}(s,\diamondsuit G)).
		\end{eqnarray*}
              }%
              where $D[s\it{\bot,\bfP(s,\cdot),t}s']$ is the policy
              that resolves nondeterminism for path $\pi'$ starting
              from $s'$ as $D$ does it for
              $s\it{\bot,\bfP(s,\cdot),t}\pi'$, i.e.
              $D(s\it{\bot,\bfP(s,\cdot),t}\pi')=\linebreak D[s\it{\bot,\bfP(s,\cdot),t}s'](\pi')$.
	\item if $s\in\PS\setminus G$, we derive
		\begin{eqnarray*}
			eT^{\min}(s,\diamondsuit G) & = & \inf_D \mathbb{E}_{s,D}(V_G) = \inf_D \int_{\paths} V_G(\pi){\Pr}_{s,D}(\mathrm{d}\pi)\\
			& = & \inf_D \sum_{s\it{\alpha,\mu,0}s'} D(s)(\alpha) \cdot \mathbb{E}_{s',D[s\it{\alpha,\mu,0}s']}(V_G).
                \end{eqnarray*}%
                where $D[s\it{\alpha,\mu,0}s']$ is the policy that
                resolves nondeterminism for path $\pi'$ starting from
                $s'$ as $D$ does it for $s\it{\alpha,\mu,0}\pi'$, i.e.
                $D(s\it{\alpha,\mu,0}\pi')=D[s\it{\alpha,\mu,0}s'](\pi')$.
                Each action $\alpha\in\Act(s)$ uniquely determines a
                distribution $\mu^s_{\alpha}$, such that the successor
                state $s'$, with $s\it{\alpha,\mu^s_{\alpha},0}s'$,
                satisfies $\mu^s_{\alpha}(s') > 0$:
		\begin{equation*}
			\alpha^\star = \argmin \left\{ \inf_{D}\sum_{s'\in S}\mu^s_{\alpha}(s')\cdot \mathbb{E}_{s',D}(V_G) \mid {\alpha\in\Act(s)} \right\}
		\end{equation*}
		Hence, all optimal policies choose $\alpha^\star$ with probability $1$, i.e. $D(s)(\alpha^\star) = 1$ and $D(s)(\beta) = 0$ for all $\beta\not=\alpha^\star$. Thus, we obtain
		\begin{eqnarray*}
			eT^{\min}(s,\diamondsuit G) & = & \inf_D \min_{s\it{\alpha}\mu^s_{\alpha}}\sum_{s'\in S}\mu^s_{\alpha}(s') \cdot \mathbb{E}_{s',D[s\it{\alpha,\mu^s_{\alpha},0}s']}(V_G)\\
			& = & \min_{s\it{\alpha}\mu^s_{\alpha}}\inf_{D} \sum_{s' \in S}\mu^s_{\alpha}(s')\cdot \mathbb{E}_{s',D[s\it{\alpha,\mu^s_{\alpha},0}s']}(V_G)\\
			& = & \min_{s\it{\alpha}\mu^s_{\alpha}}\inf_{D} \sum_{s' \in S}\mu^s_{\alpha}(s')\cdot \mathbb{E}_{s',D}(V_G)\\
			& = & \min_{s\it{\alpha}\mu^s_{\alpha}} \sum_{s' \in S}\mu^s_{\alpha}(s')\cdot eT^{\min}(s',\diamondsuit G)\\
			& = & \min_{\alpha \in \Act(s)} \sum_{s' \in S}\mu^s_{\alpha}(s')\cdot eT^{\min}(s',\diamondsuit G)\\
			& = & L(eT^{\min}(s,\diamondsuit G)).
		\end{eqnarray*}%
	\item if $s\in G$, we derive
		\[
			eT^{\min}(s,\diamondsuit G) = \inf_{D}\int_{\paths}V_G(\pi){\Pr}_{s,D}(\mathrm{d}\pi) = 0 = L(eT^{\min}(s,\diamondsuit G)).
		\]\vskip-0.8cm\qed
\end{enumerate}

\section{Proof of Theorem~\ref{thm_expected_time_reachability_reduction}}
\thmExpectedTimeReachabilityReduction*
\proof As shown in \cite{BerTsi91,deA97_thesis}, $cR^{min}(s,\diamondsuit G)$ is the unique fixpoint of the Bellman operator $L'$ defined as
\begin{equation*}
	[L'(v)](s)=\min_{\alpha\in\Act(s)} c(s,\alpha) + \sum_{s'\in S\setminus G} \bfP(s,\alpha,s')\cdot v(s') + \sum_{s'\in G}\bfP(s,\alpha,s')\cdot g(s').
\end{equation*}
We show that the Bellman operator $L$ for $\mI$ defined in Theorem \ref{thm_expected_reachability} equals $L'$ for $\mbox{\sf ssp}_{et}(\mathcal{M})$. Note that by definition $g(s)=0$ for all $s\in G$. Thus
\begin{equation*}
[L'(v)](s)=\min_{\alpha\in\Act(s)} c(s,\alpha) + \sum_{s'\in S\setminus G} \bfP(s,\alpha,s')\cdot v(s').
\end{equation*}
We distinguish three cases, $s\in\MS\setminus G, s\in\PS\setminus G$, and $s\in G$.
\renewcommand{\labelenumi}{(\roman{enumi})}
\begin{enumerate}
	%
        \item If $s\in\MS\setminus G$, then $\Act(s)=\{\bot\}$ and therefore $\min_{\alpha\in\Act(s)}c(s,\alpha)=c(s,\bot)$. Further $c(s,\bot)=\frac{1}{E(s)}$ and for all $s'\in S,\bfP(s,\bot,s')=\frac{\bfR(s,s')}{E(s)}$. Thus
	\begin{equation*}
		[L'(v)](s) = \frac{1}{E(s)} + \sum_{s'\in S} \frac{\bfR(s,s')}{E(s)}\cdot v(s') = [L(v)](s).
	\end{equation*}
	\item If $s\in\PS\setminus G$, for each action $\alpha\in\Act(s)$ and successor state $s'$, with $\bfP(s,\alpha,s') > 0$ it follows that $\bfP(s,\alpha,s')=\mu^s_{\alpha}(s')$. Further, $c(s,\alpha) = 0$ for all $\alpha\in\Act$. Thus
	{\small{
	\begin{equation*}
		[L'(v)](s) = \min_{\alpha\in\Act(s)} \sum_{s'\in S} \bfP(s,\alpha,s')\cdot v(s') = \min_{\alpha\in\Act(s)} \sum_{s'\in S} \mu^s_{\alpha}(s')\cdot v(s') = [L(v)](s).
	\end{equation*}}}
	\item If $s\in G$, then by definition $|Act(s)|=1$ with $\Act(s)=\{\bot\}$ and $\bfP(s,\bot,s)=1$ and $c(s,\bot)=0$.  Thus
	\begin{equation*}
		[L'(v)](s) = \sum_{s'\in S} \bfP(s,\alpha,s')\cdot v(s')  = 0 = [L(v)](s).
	\end{equation*}
\end{enumerate} \vspace*{-0.95cm}\qed
\section{Proof of Theorem~\ref{thm_unichain_theorem}}
First we recall the definition of weak bisimulation for MAs~\cite{EHZ10}. Therefore, we have to introduce some additional notation. A \emph{sub-distribution} $\mu$ over a set $S$ is a function $\mu\colon S\to[0,1]$ with $\sum_{s\in S}\mu(s)\leq 1$. We define $\spt(\mu)=\{s\in S \mid \mu(s)>0\}$ as the \emph{support} of $\mu$ and the probability of $S'\subseteq S$ with respect to $\mu$ as $\mu(S')=\sum_{s\in S'}\mu(s)$. Let $|\mu|:=\mu(S)$ denote the size of the sub-distribution $\mu$. If $|\mu|=1$ then $\mu$ is a full distribution. Let $\distr(S)$ and $\subdistr(S)$ denote the set of distributions and sub-distributions over $S$, respectively. We write $\dirac{s}$ for the \emph{Dirac distribution} for $s$, determined by $\dirac{s}(s) = 1$. Let $\mu$ and $\mu'$ be two sub-distributions, then $\mu'':=\mu\oplus\mu'$ is defined by $\mu''(s)=\mu(s)+\mu'(s)$, if $|\mu''|\leq 1$. Further, $\mu''$ can be split back into $\mu$ and $\mu'$, where $(\mu,\mu')$ is defined as the splitting of $\mu''$.

Next we introduce the tree notation for weak transitions. For $\sigma,\sigma'\in \mathbb{N}^*_{>0}$, let $\sigma\le\sigma'$ if there exists a (possibly empty) $\Phi\in\mathbb{N}^*_{>0}$ such that $\sigma\Phi=\sigma'$. Moreover, let $\sigma<\sigma'$ if $\sigma\le\sigma'$ and $\sigma\neq\sigma'$. A partial function $\mathcal{T}\colon \mathbb{N}^*_{>0} \to L$, which satisfies
\begin{itemize}
\item if $\sigma\leq\sigma'$ and $\sigma'\in \text{dom}(\mathcal{T})$ then $\sigma\in \text{dom}(\mathcal{T})$
\item if $\sigma i\in \text{dom}(\mathcal{T})$ for $i>1$, then also $\sigma(i-1) \in \text{dom}(\mathcal{T})$
\item $\epsilon \in \text{dom}(\mathcal{T})$
\end{itemize}
is called an (infinite) L-\emph{labelled tree}. The root of the tree $\mathcal{T}$ is called $\epsilon$ and $\sigma \in \text{dom}(\mathcal{T})$ is a node of $\mathcal{T}$. A node $\sigma$ is called a leaf of $\mathcal{T}$ if there is no $\sigma'\in \text{dom}(\mathcal{T})$ such that $\sigma<\sigma'$. We denote the set of all leaves of $\mathcal{T}$ by $\text{Leaf}_{\mathcal{T}}$ and the set of all inner nodes of $\mathcal{T}$ by $\text{Inner}_{\mathcal{T}}$.
Let $L=S\times\mathbb{R}_{\geq 0}$. A node in an $L$-labelled tree $\mathcal{T}$ is labelled by a state and the probability of reaching this node from the root node of the tree. For a node $\sigma$ we write $\text{Sta}_{\mathcal{T}}(\sigma)$ for the first component of $\mathcal{T}(\sigma)$ and $\text{Prob}_{\mathcal{T}}(\sigma)$ for the second component of $\mathcal{T}(\sigma)$.
\begin{defi}[Weak transition tree]
Let $\mathcal{M} = ( S, \Act,  \it{\, },$ $\mt{\, }, s_0)$ be an MA. A weak transition tree $\mathcal{T}$ is a $S\times\mathbb{R}_{\geq0}$-labelled tree that satisfies the following condition
\begin{enumerate}
\item $\text{Prob}_{\mathcal{T}}(\epsilon)=1$,
\item $\forall\sigma\in \text{Inner}_{\mathcal{T}}\setminus \text{Leaf}_{\mathcal{T}}:\exists\mu: \text{Sta}_{\mathcal{T}}(\sigma)\it{\,}\mu$ and\\ $\text{Prob}_{\mathcal{T}}(\sigma)\cdot\mu = \llbracket(\text{Sta}_{\mathcal{T}}(\sigma'),\text{Pro}b_{\mathcal{T}}(\sigma'))|\sigma'\in \text{Children}_{\mathcal{T}}(\sigma) \rrbracket$
\item $\sum_{\sigma\in \text{Leaf}_{\mathcal{T}}} \text{Prob}(\sigma) = 1$.
\end{enumerate}
\end{defi}
A weak transition tree $\mathcal{T}$ corresponds to a probabilistic execution fragment. It starts from $\text{Sta}_{\mathcal{T}}(\epsilon)$, and resolves the nondeterministic choices at every inner node of the tree, which represents the state in the MA it is labelled with. $\text{Prob}_{\mathcal{T}}(\sigma)$ is the probability of reaching a state $\text{Sta}_{\mathcal{T}}(\sigma)$ via immediate transitions in the MA, starting from state $\text{Sta}_{\mathcal{T}}(\epsilon)$. The distribution associated with $\mathcal{T}$, denoted $\mu_{\mathcal{T}}$, is defined as $$\mu_{\mathcal{T}}\overset{\text{def}}{=}\bigoplus_{\sigma\in\text{Leaf}_{\mathcal{T}}}\llbracket (\text{Sta}_{\mathcal{T}}(\sigma),\text{Prob}_{\mathcal{T}}(\sigma)) \rrbracket.$$ Now we can define a \emph{weak transition}: For $s\in S$ and $\mu\in \distr(S)$, let $s\rightsquigarrow \mu$ if $\mu$ is induced by some internal weak transition tree $\mathcal{T}$ with $\text{Sta}_{\mathcal{T}}(\epsilon)=s$. Let $\mu\in\distr(S)$. If for every state $s_i\in\spt(\mu)$, $s_i\rightsquigarrow\mu'_i$ for some $\mu'_i$, then we write $\mu\rightsquigarrow \bigoplus_{s_i\in\spt(\mu)}\mu(s_i)\mu'_i$.

Now a convex combination of weak transitions can be defined. Let $\mu\rightsquigarrow_C \gamma$ if there exists a finite index set $I$, and weak transitions $\mu\rightsquigarrow\gamma_i$ and a factor $c_i\in(0,1]$ for every $i\in I$, with $\sum_{i\in I} c_i=1$ and $\gamma=\bigoplus_{i\in I}c_i\gamma_i$. Let the set of splittings of immediate successor sub-distributions be defined as $\text{split}(\mu)=\{(\mu_1,\mu_2)|\exists\mu':\mu\rightsquigarrow_C\mu'\wedge\mu'=\mu_1\oplus\mu_2\}$.
\begin{defi}[Weak bisimulation]
A symmetric relation $\mathcal{R}$ on sub-distributions over $S$ is called a weak bisimulation if and only if whenever $\mu_1\mathcal{R}\mu_2$ then for all $\alpha\in\mathbb{R}\cup\{\epsilon\}:|\mu_1|=|\mu_2|$ and for all $s\in\spt(\mu_1)$ there exists $\mu^{\it{}}_2,\mu^{\Delta}_2)\in\text{split}(\mu_2)$ and
\begin{enumerate}
\item $\mu_1(s)\dirac{s} \mathcal{R}\mu_2^{\it{}}$ and $(\mu_1\ominus s) \mathcal{R}\mu_2^{\Delta}$
\item whenever $s\it{a}\mu'_1$ for some $\mu'_1$ then $\mu_2^{\it{}}\overset{a}{\rightsquigarrow_C}\mu''$ and $(\mu_1(s)\cdot\mu'_1)\mathcal{R}\mu''$
\end{enumerate}
Two sub-distributions $\mu$ and $\gamma$ are weak bisimilar, denoted $\mu\approx\gamma$, if the pair $(\mu,\gamma)$ is contained in some weak bisimulation.
\end{defi}
MA $\mI_1,\mI_2$ are weak bisimilar, denoted $\mI_1\approx\mI_2$, if their initial (Dirac) distributions are bisimilar in the direct sum.\\
\begin{lem}
For every unichain MA and stationary deterministic policy $D$, the induced stochastic process $\mI_D$ is weak bisimilar to an ergodic CTMC $C$.
\end{lem}
\proof Let $\mI_D$ be the stochastic process induced by a unichain MA $\mI$ and stationary deterministic policy $D$. As $\mI$ is unichain it directly follows that $M_D$ is strongly connected. The proof that $\mI_D$ is weakly bisimilar to a CTMC $C$ goes along the same lines as in \cite{EHKZ13} where it has been shown that the MA semantics of well-defined GSPNs is weakly bisimilar to their CTMC semantics. As the stochastic process $\mI_D$ can be considered as a 1-safe GSPN that by $D$ is well-defined, the result follows. \qed
\unichainTheorem*
\proof Let $\mI$ be a unichain MA with state space $S$ and $G\subseteq S$ a set of goal states. We consider a stationary deterministic policy $D$ on $\mI$. 
It follows that there exists an ergodic CTMC $C$ such that $\mI_D\approx C$.
Note that $G\subseteq\MS$; thus $G$ can be represented by the union of zero or more equivalence classes under $\approx$.

The long-run average for state $s\in S$ and $G\subseteq S$ is given by
 \begin{equation*}
	LRA^D(s,G)=\mathbb{E}_{s,D}(A_G)=\mathbb{E}_{s,D}\left(\lim_{t\to\infty}\frac{1}{t}\int_0^t\textbf{1}_G(\mathcal{X}_u)\dd u\right)
\end{equation*}
where $\mathcal{X}_u$ is the random variable, denoting $\pi @ u$. With the ergodic theorem from \cite{Nor97} we obtain that almost surely
 \begin{equation*}
	\frac{1}{t}\int_0^t \textbf{1}_{\{s_i\in\mathcal{X}_u\}}\dd u \to \frac{1}{m_i E(s_i)} \text{ as } t\to\infty
\end{equation*}
holds, where $m_i$ is the expected return time to state $s_i$. Therefore, in our induced ergodic CTMC, almost surely
\begin{equation}
	\label{eq:fractionTime}
	\mathbb{E}_{s_i}\left(\lim_{t\to\infty}\frac{1}{t}\int_0^t\textbf{1}_{\{s_i\}}(\mathcal{X}_u)\dd u\right) = \frac{1}{m_i\cdot E(s_i)}.
\end{equation}
Thus, almost surely the fraction of time to stay in $s_i$ in the long-run is $\frac{1}{m_i\cdot E(s_i)}$. 
Let $\mu_i$ be the probability to stay in $s_i$ in the long-run in the embedded DTMC of $C$ where $\bfP(s,s')=\frac{\bfR(s,s')}{E(s)}$. Thus $\mu\cdot\bfP=\mu$ where $\mu$ is the vector containing $\mu_i$ for all states $s_i\in S$. Given the probability of $\mu_i$ of staying in state $s_i$ the expected return time is given by
\begin{equation}
	m_i=\frac{\sum_{s_j\in S}\mu_j\cdot E(s_j)^{-1}}{\mu_i}.
	\label{eq:expRetTime}
\end{equation}
Gathering these results yields:
\begin{eqnarray*}
	LRA^D(s,G) & = & \mathbb{E}_{s,D}\left(\lim_{t\to\infty}\frac{1}{t}\int_0^t\textbf{1}_G(\mathcal{X}_u)\dd u\right)= \mathbb{E}_{s,D}\left(\lim_{t\to\infty}\frac{1}{t}\int_0^t\sum_{s_i\in G}\textbf{1}_{\{s_i\}}(\mathcal{X}_u)\dd u\right)\\
	& = & \sum_{s_i \in G} \mathbb{E}_{s,D}\left(\lim_{t\to\infty}\frac{1}{t}\int_0^t\textbf{1}_{\{s_i\}}(\mathcal{X}_u)\dd u\right) \overset{\eqref{eq:fractionTime}}{=} \sum_{s_i \in G} \frac{1}{m_i\cdot E(s_i)}\\
	& \overset{\eqref{eq:expRetTime}}{=} & \sum_{s_i \in G} \frac{\mu_i}{\sum_{s_j\in S}\mu_j\cdot E(s_j)^{-1}} \cdot \frac{1}{E(s_i)} = \frac{\sum_{s_i\in G}\mu_i\cdot E(s_i)^{-1}}{\sum_{s_j\in S}\mu_j\cdot E(s_j)^{-1}} \\
	& = & \frac{\sum_{s_i\in S}\textbf{1}_{G}(s_i)\cdot \mu_i E(s_i)^{-1}}{\sum_{s_j\in S}\mu_j\cdot E(s_j)^{-1}} = \frac{\sum_{s_i\in S}\mu_i\cdot(\textbf{1}_{G}(s_i)\cdot E(s_i)^{-1})}{\sum_{s_j\in S}\mu_j\cdot E(s_j)^{-1}}\\
	& = &  \frac{\sum_{s_i\in S}\mu_i\cdot c_1(s_i,D(s_i))}{\sum_{s_j\in S}\mu_j\cdot c_2(s_j,D(s_j))} \overset{\text{\cite{DBLP:conf/lics/Alfaro98}}}{=} \mathbb{E}_{s,D}(\mathcal{R})
\end{eqnarray*}
Thus, by definition there exists a one-to-one correspondence between the policy $D$ of $\mI$ and its corresponding MDP $\mbox{\sf mdp}(\mI)$. With the results from above this yields that $\LRA^{min}(s,G)=\inf_D\LRA^D(s,G)$ in MA $\mI$ equals $R^{min}(s,\diamondsuit G)=\inf_D\mathbb{E}_{s,D}(\mathcal{R})$ in $\mbox{\sf mdp}(\mI)$.\qed
\section{Proof of Lemma~\ref{lem:lemMEC}}
\lemMEC*
\proof (sketch). By the limit in the long-run ratio definition of $\mathcal{R}$ it follows that for every $i\geq 0$ the prefix of $\pi$ up to $i$ does not matter. Thus, $\mathcal{R}(\pi)=\mathcal{R}(\pi_i)$ where $\pi_i$ denotes the path $\pi$ from the $i$-th position onwards. Therefore, given a policy $D$, inducing a multichain on maximal end component $\mI$, we can construct a unichain policy $D'$ as follows: Let $D'$ fixes the recurrent class $S'$ of $\mI$ with the minimal value induced by $D$ (in case of the maximal long-run ratio, the maximal value respectively). For states outside of $S'$, $D'$ is a policy that reaches $S'$ with probability $1$. \qed
\section{Proof of Theorem~\ref{lem:LRA_SSP}}
\lemLRASSP*
\proof (sketch). Let $\mI$ be a finite MA with maximal end components $\{\mI_1,\ldots,\mI_k\}$, $G\subseteq S$ a set of goal states, and $\pi\in\Paths(\mI)$ an infinite path in $\mI$. 
For all policies $D$, each path $\pi$ can be partitioned into finite and infinite paths of the form
\begin{eqnarray*}
\pi_{s_0s}& = &s_0\it{\alpha_0,\mu_0,t_0}s_1 \it{\alpha_1,\mu_1,t_1} \ldots \it{\alpha_n,\mu_n,t_n} s, \text{ and}\\
\pi_{s}^{\omega}& = &s\it{\alpha_s,\mu_s,t_s} \ldots \it{\alpha_i,\mu_i,t_i} s \ldots
\end{eqnarray*}
where $\pi_{s_0s}$ is the path starting in initial state $s_0$ and ends in $s\in\mI_i$ for some $0<i\leq k$. Further, all states on path $\pi^{\omega}_s$ belong to maximal end component $\mI_i$. Note, that a state on path $\pi_{s_0s}$ can be part of another maximal end component $\mathcal{M}_j$ (as in Example~\ref{ex:append}). Hence, it is not sufficient to only check if eventually a MEC is reached, as done in the corresponding theorem for IMCs in~\cite{DBLP:conf/nfm/GuckHKN12}. Thus, the minimal LRA will be obtained when the LRA in each MEC $\mI_i$ is minimal and the combined LRA of all MECs is minimal according to their persistence under policy $D$. \qed
\section{Proof of Theorem~\ref{thm:LRA_SSP}}
\thmLRASSP*
\proof
Let $\pi$ be an infinite path in the MDP $\mbox{\sf ssp}_{lra}(\mathcal{M})$ such that $\pi[i_Q]$ is the first visit of a state in $Q$ along $\pi$, i.e., for all $j < i_Q$, $\pi[j] \not\in Q$ and $\pi[i_Q] \in Q$.  Similarly, we define $i_q$ for a single state $q$. We define random variable $C_Q:\paths\rightarrow\nnreal$ by $C_Q(\pi)=g(\pi[i_Q])$. Note that $D\in\GMS$.
\begin{eqnarray*}
	cR^{min}(s_0,\diamondsuit Q)
	&=& \inf_{D} \, \mathbb{E}_{s_0,D}(C_Q) \\
	&=& \inf_D\sum_{\pi\in\paths} C_Q(\pi) \cdot {\Pr}_{s_0,D}(\pi)=\inf_D\sum_{\pi\in\paths}\sum_{i=1}^{k}C_{\{q_i\}}(\pi) \cdot {\Pr}_{s_0,D}(\pi)\\
	&=&\inf_D\sum_{i=1}^{k}\sum_{\pi\in\paths}C_{\{q_i\}}(\pi) \cdot {\Pr}_{s_0,D}(\pi)= \inf_D \sum_{i=1}^{k} \LRA_{i}^{min}(G)\cdot {\Pr}_{s_0,D}(\diamondsuit \{q_i\})\\
	&\overset{(*)}{=}  & \inf_D \sum_{i=1}^{k} \LRA_{i}^{min}(G)\cdot {\Pr}_{s_0,D}(\diamondsuit\G S_i)\ \ = \ \ \LRA^{min}(s_0,G).
\end{eqnarray*}
Observe that in step $(*)$ we use the transformation from Definition~\ref{defSSP} in reverse. Hence, if ${\Pr}_{s,D}(\diamondsuit q_i)>0$, we eventually reach the maximal end component $\mathcal{M}_i$ and always stay in it. Otherwise ${\Pr}_{s,D}(\diamondsuit q_i)=0$ and policy $D$ chooses an action such that we leave $\mathcal{M}_i$ or never even visit $\mathcal{M}_i$.\qed
\section{Proof of Theorem \ref{thm:tbr}}
\def\hash{{\scriptstyle\#}} Let MA \defma, $G \subseteq S$ and time
interval $I=[0,b]\in\mathcal{Q}$ with $b\ge 0$. Let $\lambda = \max_{s
  \in \MaS}E(s)$ be the largest exit rate of any Markovian state and
$\delta\in\mathbb{R}_{>0}$ be the discretisation step, chosen such
that $b=k_b\delta$ for some $k_b \in \mathbb{N}$. We recall the
definition of $\diamondsuit^IG$ as the set of all paths that reach
some goal state in $G$ within interval $I$. Let random variable
$\hash_{J}\colon\paths\rightarrow\mathbb{N}$, where $J \in
\mathcal{Q}$ is a time interval. Intuitively $\hash_J$ counts the
number of Markovian jumps happened inside interval $J$. For example
$\hash_{[0,\delta]}=1$ denotes the set of paths having exactly one
Markovian transition in the first $\delta$ time units. Random vector
$\hash^{I,\delta}\colon\paths\rightarrow\mathbb{N}^{k_b}$ with $I$,
$\delta$ and $k_b$ as explained before, is defined as the vector of
$k_b$ elements, each counting Markovian jumps occurred in the
corresponding chunk of length $\delta$, i.e.
$\hash^{I,\delta}=\left(\hash_{[0,\delta)}, \dotsc,
  \hash_{[(k_b-2)\delta,(k_b-1)\delta)},
  \hash_{[(k_b-1)\delta,b]}\right)^{\text{T}}$.  Moreover, let
$\|\cdot\|_\infty$ denote the maximum norm, which takes the maximum
over the absolute value of the elements of the given vector.

\begin{lem}\label{lemma:dma}
  Let $\MAM_{\delta}$ be the dMA induced by \MAM with respect to
  discretisation constant $\delta$. Then for all $s \in S$:
  \begin{equation*}
    p^{\MAM_{\delta}}_{\max}(s, \rchgls{[0,k_b]})=\sup_{D \in \GMS} {\Pr}_{s,D}(\rchgls{I} \mid \left \|\hash^{I,\delta}  \right \|_{\infty} < 2).
  \end{equation*}
\end{lem}
\proof As discussed in Section~\ref{section:timed}, paths of
$\MAM_{\delta}$ are essentially the paths from \MAM that carry only
zero or one Markovian transitions in each discretisation step
$\delta$. Hence, for computing reachability probabilities in step
interval $[0,k_b]$ in $\MAM_{\delta}$, it is enough to consider paths
in \MAM with at most one Markovian jumps in each $\delta$ time
units. This set
  is described by $\left \| \hash^{I,\delta} \right \|_{\infty} < 2$.
\qed

\begin{lem}\label{lemma:lb1} For all $s \in S$ and $D \in \GMS$ in
  \MAM: ${\Pr}_{s,D}(\rchgls{I} \mid \hash_{[0,\delta]}<2) \le
  {\Pr}_{s,D}(\rchgls{I})$.
\end{lem}

\proof
  We assume $b>0$, since for $b=0$, ${\Pr}_{s,D}(\rchgls{I} \mid
  \hash_{[0,\delta]}<2)={\Pr}_{s,D}(\rchgls{I})$. We have
\begin{flalign}
 {\Pr}_{s,D}(\rchgls{I}) \nonumber &={} {\Pr}_{s,D}(\rchgls{I} \cap \hash_{[0,\delta]} > 0) + {\Pr}_{s,D}(\rchgls{I} \cap \hash_{[0,\delta]} = 0)& \nonumber\\
 &={} {\Pr}_{s,D}(\rchgls{I} \cap \hash_{[0,\delta]} > 0) + {\Pr}_{s,D}(\rchgls{I} \mid \hash_{[0,\delta]} = 0)\cdot{\Pr}_{s,D}(\hash_{[0,\delta]} = 0).\hspace{-10pt}& \label{eq:tbr}
\end{flalign}
On the other hand we have
\begin{flalign}
 {\Pr}_{s,D}(\rchgls{I} \mid \hash_{[0,\delta]}<2)  ={}&{\Pr}_{s,D}(\rchgls{I} \mid \hash_{[0,\delta]} <2, \hash_{[0,\delta]} =1 )\cdot{\Pr}_{s,D}( \hash_{[0,\delta]} =1 \mid \hash_{[0,\delta]} <2)& \nonumber \\ &+  {\Pr}_{s,D}(\rchgls{I} \mid \hash_{[0,\delta]} <2, \hash_{[0,\delta]}=0)\cdot{\Pr}_{s,D}( \hash_{[0,\delta]} = 0 \mid \hash_{[0,\delta]} <2).& \label{eq:tbrapprox}
\end{flalign}
We distinguish between two cases:
\begin{enumerate}
\item $s \in \MS \setminus G$: In this case, Eq.~\eqref{eq:tbr} gives
  \begin{flalign}
    &{\Pr}_{s,D}(\rchgls{I}) = \int_0^{\delta} E(s) \ee^{-E(s)t} \sum_{s' \in S} \bfP (s,s') {\Pr}_{s',D}(\rchgls{I \ominus t})\dd t + {\Pr}_{s,D}(\rchgls{I \ominus
      \delta})\ee^{-E(s)\delta}.& \label{eq:tbrms}
  \end{flalign}
  and for Eq.~\eqref{eq:tbrapprox} we have
  \begin{flalign}
    {\Pr}_{s,D}(\rchgls{I}\mid \hash_{[0,\delta]}<2) = {} &\int_0^{\delta} E(s) \ee^{-E(s)t} \sum_{s' \in S} \bfP (s,s') {\Pr}_{s',D}(\rchgls{I \ominus\delta})\dd t \nonumber &\\
    &+ {\Pr}_{s,D}(\rchgls{I \ominus
      \delta})\ee^{-E(s)\delta}.&\label{eq:tbrapproxms}
  \end{flalign}
  Since ${\Pr}_{s,D}(\rchgls{I \ominus t})$ is monotonically
  decreasing in $t$, we have ${\Pr}_{s,D}(\rchgls{I \ominus \delta})
  \le {\Pr}_{s,D}(\rchgls{I \ominus t}), \; t \le \delta$. Putting
  this in Eq.~\eqref{eq:tbrms} and~\eqref{eq:tbrapproxms} leads to
  \begin{equation*}
    {\Pr}_{s,D}(\rchgls{I} \mid \hash_{[0,\delta]}<2) \le {\Pr}_{s,D}(\rchgls{I}).
  \end{equation*}
\item $s \in \PS \setminus G$: From the law of total probability, we
  split time bounded reachability into two parts. First we compute the
  probability to reach the set of Markovian states from $s$ by only
  taking probabilistic transitions in zero time, and then we quantify
  the probability to reach some goal state in $G$ from Markovian
  states inside interval~$I$. Therefore:
\begin{flalign*}
{\Pr}_{s,D}(\rchgls{I}) = {} & \sum_{s' \in \scriptsize{\MS}} {\Pr}_{s,D}(\diamondsuit^{[0,0]}\{s'\}) {\Pr}_{s',D}(\rchgls{I}) \\
        \overset{(*)}{\ge}{} & \sum_{s' \in \scriptsize{\MS}} {\Pr}_{s,D}(\diamondsuit^{[0,0]}\{s'\}) {\Pr}_{s',D}(\rchgls{I}\mid \hash_{[0,\delta]} < 2 ) \\
                        = {} & {\Pr}_{s,D}(\rchgls{I}\mid \hash_{[0,\delta]} < 2),
\end{flalign*}
where $(*)$ follows from case (i) above.\qed
\end{enumerate}

\begin{lem}\label{lemma:lb2} For all $s \in S \setminus G$ and $D \in \GMS$ in \MAM:
  \begin{equation*}
  {\Pr}_{s,D}(\rchgls{I} \mid \left \|\hash^{I,\delta}  \right \|_{\infty} < 2) \le {\Pr}_{s,D}(\rchgls{I} \mid \hash_{[0,\delta]}<2).
  \end{equation*}
\end{lem}
\proof
  The lemma holds for $b=0$, since in this case,
  ${\Pr}_{s,D}(\rchgls{I} \mid \left \|\hash^{I,\delta} \right
  \|_{\infty} < 2) = {\Pr}_{s,D}(\rchgls{I} \mid
  \hash_{[0,\delta]}<2)$. For $b>0$, we decompose
  ${\Pr}_{s,D}(\rchgls{I} \mid \left \|\hash^{I,\delta} \right
  \|_{\infty} < 2)$ as Eq.~\eqref{eq:tbrapprox} into:
  \begin{multline}
    {\Pr}_{s,D}(\rchgls{I} \mid \left \|\hash^{I,\delta} \right
    \|_{\infty} < 2, \hash_{[0,\delta]} =1 )\cdot {\Pr}_{s,D}(
    \hash_{[0,\delta]} =1 \mid \left \|\hash^{I,\delta} \right
    \|_{\infty} < 2)  \\ + {\Pr}_{s,D}(\rchgls{I} \mid \left \|\hash^{I,\delta} \right
    \|_{\infty} < 2, \hash_{[0,\delta]} =0 )\cdot {\Pr}_{s,D}(
    \hash_{[0,\delta]} = 0 \mid \left \|\hash^{I,\delta} \right
    \|_{\infty} < 2).\label{eq:tbrdig}
  \end{multline}
  Now we prove the lemma by induction over $k_b$.
  \begin{itemize}
  \item $k_b=1$: This case holds because interval $I=[0,\delta]$ contains one discretisation step and then ${\Pr}_{s,D}(\rchgls{I} \mid \left \|\hash^{I,\delta}  \right \|_{\infty} < 2) = {\Pr}_{s,D}(\rchgls{I} \mid \hash_{[0,\delta]}<2)$.
  \item $k_b - 1 \leadsto k_b$: Let $I$ be $[0,b]$ and assume the
    lemma holds for interval $[0,(k_b -1)\delta]$
    (i.e. $I\ominus\delta$):
  \begin{equation}
    {\Pr}_{s,D}(\rchgls{I\ominus\delta} \mid \left \|\hash^{I\ominus\delta,\delta}  \right \|_{\infty} < 2) \le {\Pr}_{s,D}(\rchgls{I\ominus\delta} \mid \hash_{[0,\delta]}<2).\label{eq:ih}
  \end{equation}
  In order to show that the lemma holds for $I$, we distinguish
  between two cases:
  \begin{enumerate}
  \item $s \in \MS \setminus G$: From Eq.~\eqref{eq:tbrapproxms} we have:
    \begin{equation}
      {\Pr}_{s,D}(\rchgls{I}\mid \hash_{[0,\delta]}<2) =  \sum_{s' \in S} \bfP (s,s') {\Pr}_{s',D}(\rchgls{I \ominus\delta})(1-\ee^{-E(s)\delta})
      + {\Pr}_{s,D}(\rchgls{I \ominus
        \delta})\ee^{-E(s)\delta}.\label{eq:tbrapproxmssim}
    \end{equation}
    Similarly from Eq.~\eqref{eq:tbrdig} we have:
    \begin{flalign*}
      {\Pr}_{s,D}(\rchgls{I}\mid \left \| \hash^{I,\delta}  \right \|_{\infty} < 2)= {} &\sum_{s' \in S} \bfP (s,s') {\Pr}_{s',D}(\rchgls{I \ominus\delta} \mid \left \|\hash^{I\ominus\delta,\delta}  \right \|_{\infty} < 2)(1-\ee^{-E(s)\delta}) &\\
      &+ {\Pr}_{s,D}(\rchgls{I \ominus
        \delta}\mid \left \|\hash^{I\ominus\delta,\delta}  \right \|_{\infty} < 2)\ee^{-E(s)\delta}&\\
      \overset{\eqref{eq:ih}}{\le} {} & \sum_{s' \in S} \bfP (s,s') {\Pr}_{s',D}(\rchgls{I \ominus\delta})(1-\ee^{-E(s)\delta})& \\&+ {\Pr}_{s,D}(\rchgls{I \ominus
        \delta})\ee^{-E(s)\delta}&\\
      \overset{\eqref{eq:tbrapproxmssim}}{=} {} &
      {\Pr}_{s,D}(\rchgls{I}\mid \hash_{[0,\delta]}<2)&
    \end{flalign*}
  \item $s \in \PS \setminus G$: This case utilises the previously
    discussed idea of splitting paths using the law of total
    probabilities into two parts. The first part contains the set of
    paths that reach Markovian states from $s$ in zero time using
    probabilistic transitions, while the second includes paths
    reaching some state in $G$ from Markovian states. Hence:
    \begin{flalign*}
      {\Pr}_{s,D}(\rchgls{I}\mid \left \|\hash^{I,\delta}  \right \|_{\infty} < 2) = {} & \sum_{s' \in {\scriptsize\MS}} {\Pr}_{s,D}(\diamondsuit^{[0,0]}\{s'\}) {\Pr}_{s',D}(\rchgls{I}\mid \left \|\hash^{I,\delta}  \right \|_{\infty} < 2)& \\
      \overset{(*)}{\le}{} & \sum_{s' \in {\scriptsize\MS}} {\Pr}_{s,D}(\diamondsuit^{[0,0]}\{s'\}) {\Pr}_{s',D}(\rchgls{I}\mid \hash_{[0,\delta]} < 2 )& \\
      = {} & {\Pr}_{s,D}(\rchgls{I}\mid \hash_{[0,\delta]} < 2 ),
    \end{flalign*}
    where $(*)$ follows from case (i) above.\qed
  \end{enumerate}
  \end{itemize}

\begin{lem}\label{lemma:lb} For all $s \in S \setminus G$:
  $p^{\MAM_{\delta}}_{\max}(s,\rchgls{[0,k_b]}) \le p^{\MAM}_{\max}(s,\rchgls{I})$.
\end{lem}
\proof
\begin{flalign*}
 p^{\MAM_{\delta}}_{\max}(s, \rchgls{[0,k_b]})={}& \sup_{D \in \GMS} {\Pr}_{s,D}(\rchgls{I} \mid \left \|\hash^{I,\delta}  \right \|_{\infty} < 2) &&(\text{Lemma~\ref{lemma:dma}})&\\
\le{} & \sup_{D \in \GMS} {\Pr}_{s,D}(\rchgls{I} \mid  \hash_{[0,\delta]} < 2) &&(\text{Lemma~\ref{lemma:lb2}})&\\
\le{} & \sup_{D \in \GMS} {\Pr}_{s,D}(\rchgls{I}) =  p^{\MAM}_{\max}(s,\rchgls{I}). &&(\text{Lemma~\ref{lemma:lb1}})&
\end{flalign*}\qed

\begin{lem}\label{lemma:ub} For all $s \in S \setminus G$:
\begin{equation*}
  p^{\MAM}_{\max}(s,\rchgls{I}) \le p^{\MAM_{\delta}}_{\max}(s,\rchgls{[0,k_b]}) + 1 - \ee^{-\lambda b}(1 + \lambda\delta)^{k_b}.
\end{equation*}
\end{lem}

\proof
  \begin{flalign*}
    p^{\MAM}_{\max}(s,\rchgls{I}) = {}& \sup_{D \in \GMS} {\Pr}_{s,D}(\rchgls{I})\\
    = {}& \sup_{D \in \GMS} \Big({\Pr}_{s,D}(\rchgls{I} \cap \left \| \hash^{I,\delta} \right \|_{\infty} < 2 ) + {\Pr}_{s,D}(\rchgls{I} \cap \left \| \hash^{I,\delta} \right \|_{\infty} \ge 2)\Big) &&\\
    \le {}& \sup_{D \in \GMS} {\Pr}_{s,D}(\rchgls{I} \cap \left \| \hash^{I,\delta} \right \|_{\infty} < 2 ) + \sup_{D \in \GMS}{\Pr}_{s,D}(\rchgls{I} \cap \left \| \hash^{I,\delta} \right \|_{\infty} \ge 2)&& \\
    \le {}& \sup_{D \in \GMS} {\Pr}_{s,D}(\rchgls{I} \mid \left \| \hash^{I,\delta} \right \|_{\infty} < 2 ) + \sup_{D \in \GMS}{\Pr}_{s,D}(\rchgls{I} \cap \left \| \hash^{I,\delta} \right \|_{\infty} \ge 2) &&\\
    \overset{(\dag)}{=} {}& p^{\MAM_{\delta}}_{\max}(s,\rchgls{[0,k_b]}) + \sup_{D \in \GMS}{\Pr}_{s,D}(\rchgls{I} \cap \left \| \hash^{I,\delta} \right \|_{\infty} \ge 2) &&\\
                                 \le {}& p^{\MAM_{\delta}}_{\max}(s,\rchgls{[0,k_b]}) + \sup_{D \in \GMS}{\Pr}_{s,D}(\left \| \hash^{I,\delta} \right \|_{\infty} \ge 2), &&
  \end{flalign*}
  where $(\dag)$ follows from Lemma~\ref{lemma:dma}.

  It remains to find an upper bound for $\sup_{D \in
    \GMS}{\Pr}_{s,D}(\left \| \hash^{I,\delta} \right \|_{\infty} \ge
  2)$ which is the maximum probability to have more than one Markovian
  jump in at least one time step among $k_b$ time step(s) of length
  $\delta$.  Due to the independence of the number of Markovian jumps
  in discretisation steps, this probability can be upper bounded by
  $k_b$ independent Poisson processes, all parametrised with the
  maximum exit rate exhibited in \MAM. In each Poisson process the
  probability of at most one Markovian jump in one discretisation step
  is $\ee^{- \lambda \delta}(1 + \lambda \delta)$, therefore the
  probability of a violation of this assumption in at least one
  discretisation step is $1 - \ee^{- \lambda b}\big(1+ \lambda
  \delta\big)^{k_b}$. Hence
  \begin{align*}
    p^{\MAM}_{\max}(s,\rchgls{I}) \le {}& p^{\MAM_{\delta}}_{\max}(s,\rchgls{[0,k_b]}) + \sup_{D \in \GMS}{\Pr}_{s,D}(\left \| \hash^{I,\delta} \right \|_{\infty} \ge 2)& \\ \le {}& p^{\MAM_{\delta}}_{\max}(s,\rchgls{[0,k_b]}) + 1 - \ee^{- \lambda b}\big(1+ \lambda \delta\big)^{k_b}.
  \end{align*}\vskip-0.6cm\qed

\thmTBR*
\proof
  For $s\in G$ we have that $p^{\mathcal{M}_{\delta}}_{\max}(s,
  \diamondsuit^{[0,k_b]} \, G) = p^{\mathcal{M}}_{\max}(s,
  \diamondsuit^{[0,b]} \, G)=1$. For $s \in S \setminus G$, the theorem follows
  from Lemma~\ref{lemma:lb} and~\ref{lemma:ub}.\qed

\vspace{-40 pt}
\end{document}